\documentclass{tdp}
\usepackage{amsmath,amsfonts}
\usepackage{amsthm}
\usepackage{algorithmic}
\usepackage{algorithm}
\usepackage{array}
\usepackage[caption=false,font=normalsize,labelfont=sf,textfont=sf]{subfig}
\usepackage{textcomp}
\usepackage{stfloats}
\usepackage{url}
\usepackage{verbatim}
\usepackage{graphicx}
\usepackage{color}
\usepackage{enumitem}
\usepackage{multirow}

\newcommand{\nats}{\mathbb{N}}
\newcommand{\nnints}{\mathbb{Z}_{\ge0}}
\newcommand{\reals}{\mathbb{R}}
\newcommand{\nnreals}{\mathbb{R}_{\ge0}}

\renewcommand{\epsilon}{\varepsilon}

\newcommand{\calG}{\mathcal{G}}

\newcommand{\calN}{\mathcal{N}}
\newcommand{\calR}{\mathcal{R}\hspace{0.1mm}}

\newcommand{\bmA}{\mathbf{A}}

\newcommand{\bma}{\mathbf{a}}

\newcommand{\ta}{\tilde{a}}
\newcommand{\td}{\tilde{d}}

\newcommand{\tbma}{\tilde{\bma}}
\newcommand{\tbmA}{\tilde{\bmA}}

\newcommand{\Agg}{\textsf{AGGREGATE}}
\newcommand{\Com}{\textsf{COMBINE}}
\newcommand{\Read}{\textsf{READOUT}}

\newcommand{\LocalLap}{\text{LocalLap}}
\newcommand{\NPPartial}{\text{NonPriv-Part}}
\newcommand{\NPFull}{\text{NonPriv-Full}}
\newcommand{\Poison}{\text{Poison}}
\newcommand{\Defense}{\text{Defense}}

\newcommand{\Lap}{\textrm{Lap}}

\newcommand{\colorB}[1]{\textcolor{black}{#1}}

\newif\ifconferenceon\conferenceontrue
\ifconferenceon
\newcommand{\conference}[1]{#1}
\newcommand{\arxiv}[1]{}
\else
\newcommand{\conference}[1]{}
\newcommand{\arxiv}[1]{#1}
\usepackage{balance}
\fi

\begin{document}

\title{Degree-Preserving Randomized Response for Graph Neural Networks under Local Differential Privacy}
\author{Seira Hidano$^{*}$, Takao Murakami$^{**,***}$}
\address{$^{*}$KDDI Research, Inc., 2-1-15 Ohara, Fujimino, Saitama, 356-8502, Japan.\\
  $^{**}$ISM, 10-3 Midori-cho, Tachikawa, Tokyo, 190-8562, Japan.\\
  $^{***}$AIST, 2-4-7 Aomi, Koto-ku, Tokyo, 135-0064, Japan.\\
  E-mail: {\small \tt{se-hidano@kddi.com}}, {\small \tt{tmura@ism.ac.jp}}
}
\thanks{The first and second authors made equal contribution.} 

\TDPRunningAuthors{Seira Hidano, Takao Murakami}
\TDPRunningTitle{Degree-Preserving Randomized Response for Graph Neural Networks ...}
\TDPThisVolume{17}
\TDPThisYear{2024}
\TDPFirstPageNumber{89}

\maketitle

\begin{abstract}
Differentially private GNNs (Graph Neural Networks) have been recently studied to provide high accuracy in various tasks on graph data while strongly protecting user privacy. 
In particular, a recent study proposes an algorithm to protect each user's feature vector in \colorB{an attributed graph, which includes feature vectors along with node IDs and edges,} 
with LDP (Local Differential Privacy), a strong privacy notion without a trusted third party. 
However, this algorithm does not protect edges (friendships) in a social graph, hence cannot 
protect user privacy in \colorB{unattributed graphs, which include only node IDs and edges}. 
How to 
provide strong privacy 
with high accuracy 
in 
unattributed graphs 
remains open. 
In this paper, 
we propose a novel LDP algorithm called the \textit{DPRR (Degree-Preserving Randomized Response)} 
to provide LDP for edges in GNNs. 
Our DPRR preserves each user's degree hence a graph structure while providing edge LDP. 
Technically, 
our DPRR uses Warner's RR (Randomized Response) and \textit{strategic} edge sampling, where each user's sampling probability is automatically tuned using the Laplacian mechanism to preserve the degree information under edge LDP. 
We also propose a privacy budget allocation method to make the noise in both Warner's RR and the Laplacian mechanism small. 
We focus on graph classification as a task of GNNs and evaluate the DPRR using 
three 
social graph datasets. 
Our experimental results show that the DPRR significantly outperforms 
three 
baselines 
and 
provides accuracy close to a non-private algorithm in all datasets with a reasonable privacy budget, e.g., $\epsilon=1$. 
\colorB{Finally, we introduce data poisoning attacks to our DPRR and a defense against the attacks. 
We evaluate them using the three social graph datasets and discuss the experimental results.}
\end{abstract}


\section{Introduction}
\label{sec:intro}
Many real-world data are represented as graphs, e.g., social networks, communication networks, 
and epidemiological networks. 
GNNs (Graph Neural Networks) 
\cite{GNN_Ma} 
have recently attracted much attention because they provide state-of-the-art performance in various tasks on graph data, such as node classification \cite{Kipf_ICLR17}, graph classification \cite{Xu_ICLR19}, and community detection \cite{Chen_ICLR19}. 
However, the use of graph data raises serious privacy concerns, as it may reveal some sensitive data, such as sensitive edges (i.e., friendships in social graphs) \cite{He_USENIX21}. 

DP (Differential Privacy) 
\cite{DP} 
has been widely studied to protect user privacy strongly and is recognized as a gold standard for data privacy. 
DP provides user privacy against adversaries with any background knowledge when a parameter called the privacy budget $\epsilon$ is small, e.g., $\epsilon \leq 1$ 
or $2$ \cite{Hoshino_JJSDS20,DP_Li}. 
According to the underlying architecture, DP can be divided into two types: \textit{centralized DP} and \textit{LDP (Local DP)}. 
Centralized DP assumes a centralized model in which a trusted server has the personal data of all users and releases obfuscated versions of statistics or machine learning models. 
However, this model has a risk that the personal data of all users are leaked from the server by illegal access \cite{data_breach2021} or internal fraud \cite{CambridgeAnalytica}. 
In contrast, LDP assumes a local model where a user obfuscates her personal data by herself; i.e., it does not assume a trusted third party. 
Thus, LDP does not suffer from the data leakage issue explained above, and therefore it has been widely adopted in both the academic field 
\cite{Acharya_AISTATS19,Bassily_STOC15,Qin_CCS16,Wang_USENIX17} 
and industry 
\cite{Ding_NIPS17,Erlingsson_CCS14}. 

DP has been recently applied to GNNs \cite{Jin_arXiv22,Lin_TIFS22,Mueller_arXiv22,Olatunji_arXiv21,Sajadmanesh_CCS21,Sajadmanesh_USENIX23,Wu_arXiv21_2,Wu_arXiv21,Zhang_IJCAI21}, and most of them adopt centralized DP.  
However, they suffer from the data breach issue explained above. 
Moreover, they cannot be applied to \textit{decentralized SNSs (Social Networking Services)} \cite{Paul_CN14} such as diaspora* \cite{Diaspora} and Mastodon \cite{Mastodon}. 
In decentralized SNSs, the entire graph is distributed across many servers, and each server has only the data of the users who choose it. 
Because centralized DP algorithms take the entire graph as input, they cannot be applied to the decentralized SNSs. 
We can also consider 
\textit{fully decentralized SNSs}, where a server does not have any edge. 
For example, we can consider applications where each user sends noisy versions of her friends to the server, which then calculates some graph statistics 
\cite{Imola_USENIX21,Imola_USENIX22,Ye_TKDE21} 
or generates synthetic graphs \cite{Qin_CCS17}. 
Centralized DP cannot be applied to these applications either. 

LDP can be applied to these decentralized SNSs and does not suffer from data leakage. 
Sajadmanesh and Gatica-Perez~\cite{Sajadmanesh_CCS21} apply LDP to each user's feature vector 
(attribute values such as age, gender, city, and personal website) 
in GNNs. 
They focus on node classification and show that their algorithm provides high classification accuracy with a small privacy budget $\epsilon$ in LDP, e.g., $\epsilon=1$. 

However, they focus on hiding feature vectors and do not hide edges (friendships). 
In practice, edges include highly sensitive information, i.e., sensitive friendships. 
In addition, many social graphs are \textit{unattributed} graphs~\cite{Yanardag_KDD15}, 
which does not include feature vectors and 
include only node IDs and edges. 
Unfortunately, the algorithm in \cite{Sajadmanesh_CCS21} 
cannot be used to protect user privacy 
in unattributed graphs. 

To fill this gap, we focus on 
\textit{LDP for edges in an unattributed graph}, 
i.e., DP in a scenario where each user obfuscates her edges (neighbor list or friend list) by herself. 
We also focus on \textit{graph classification} 
as a task of GNNs (see Section~\ref{sub:system} for details) 
because a state-of-the-art GNN \cite{Xu_ICLR19} provides high accuracy for unattributed social graphs in this task. 
A recent study \cite{Zhang_USENIX22} also shows that a lot of private information can be inferred from the output of GNNs for graph classification. 
Therefore, LDP algorithms for edges in graph classification are urgently needed. 

We first show that 
Warner's RR (Randomized Response) \cite{Warner_JASA65}, which is widely used to provide LDP for edges in applications other than GNNs 
\cite{Imola_USENIX21,Imola_USENIX22,Qin_CCS17,Ye_TKDE21}, 
is insufficient for GNNs. 
Specifically, the RR flips each 1 (edge) or 0 (no edge) with some probability. 
The flipping probability is close to $0.5$ when $\epsilon$ is close to $0$. 
Consequently, it makes a sparse graph dense and destroys a graph structure\footnote{\colorB{Note that some studies \cite{Imola_USENIX22,Salas_MDAI23} propose a variant of Warner's RR that makes a graph sparse. We explain the difference between our DPRR and \cite{Imola_USENIX22,Salas_MDAI23} at the end of Section~\ref{sec:intro}.}}. 
In particular, 
GNNs use a neighborhood aggregation (or message passing) strategy, which updates each node's feature vector by aggregating feature vectors of adjacent nodes. 
In an unattributed graph, a constant value is often used as a feature vector \cite{Errica_ICLR20,Xu_ICLR19}, and in this case, the sum of adjacent feature vectors is a degree. 
Moreover, 
the degree distribution correlates with the graph type, 
as shown in our experiments. 
Thus, each user's degree 
is especially important in GNNs. 
The RR does not preserve the degree information and therefore does not provide high accuracy in GNNs for sparse graphs. 

In addition, 
the RR does not provide high accuracy even in a \textit{customized privacy setting} (as in Facebook \cite{Facebook_privacy}) where some 
users hide their neighbor lists and other 
users reveal their neighbor lists. 
We refer to the former users as \textit{private users} and the latter as \textit{non-private users}.  
The RR makes the neighbor lists of the 
private 
users dense. 
Consequently, it destroys the graph structure and ruins the neighborhood aggregation for non-private users. 
Thus, 
the accuracy is hardly increased with an increase in the non-private users, as shown in our experiments. 
Moreover, the RR has large time and space complexity and is impractical for large-scale graphs; e.g., the memory size is $1$ TB in the Orkut social network \cite{Yang_ICDM12} with 
three million users. 

To address these issues, we propose a novel LDP algorithm, which we call the \textit{DPRR (Degree-Preserving Randomized Response)}. 
\colorB{Our DPRR uses edge sampling \cite{Bera_KDD20,Eden_FOCS15,Imola_USENIX22,Wu_TKDE16} preceded with Warner's RR, and works as follows. 
First, it adds the Laplacian noise to each user's degree to provide edge LDP \cite{Qin_CCS17}. 
Then, it tunes the sampling probability based on the noisy degree. 
Finally, it applies Warner's RR and edge sampling so that each user's degree information is preserved under edge LDP. 
We use two privacy budgets, $\epsilon_1$ and $\epsilon_2$, for the Laplacian mechanism and Warner's RR, respectively. 
By the (general) sequential composition \cite{DP_Li}, our DPRR provides $(\epsilon_1 + \epsilon_2)$-edge LDP.} 

\colorB{Since our DPRR preserves each user's degree information,} 
it is suitable for the neighborhood aggregation strategy in GNNs. 
It also works very well in the customized privacy setting -- the accuracy is rapidly increased with an increase in non-private users. 
We show that our DPRR significantly outperforms Warner's RR, especially in the customized setting. 
Moreover, 
we show that the DPRR is much more efficient than the RR in that the DPRR needs much less time for both training and classification and much less memory; e.g., the memory size is about $30$ MB even in the Orkut social network explained above. 

We also compare our DPRR with 
two 
other private baselines: (i) a local model version of LAPGRAPH (Laplace Mechanism for Graphs) \cite{Wu_arXiv21} 
and 
(ii) 
an algorithm that 
discards neighbor lists of private users and uses a graph composed of only non-private users. 
We denote the former and latter baselines by \LocalLap{} and \NPPartial{}, respectively. 
\colorB{The latter baseline, \NPPartial{}, is an algorithm that has not been studied in the literature. 
\NPPartial{} uses only the information of non-private users. 
We introduce this baseline to show the effectiveness of using the information of both private and non-private users.} 
We show that the DPRR significantly outperforms 
\LocalLap{} and \NPPartial{} 
and 
provides high accuracy 
with a reasonable privacy budget, e.g., $\epsilon=1$. 

\colorB{Note that we assume an \textit{honest-but-curious} setting, as with most of the existing work on LDP (e.g., \cite{Acharya_AISTATS19,Bassily_STOC15,Erlingsson_CCS14,Imola_USENIX21,Imola_USENIX22,Qin_CCS16,Qin_CCS17,Wang_USENIX17,Ye_TKDE21}). 
That is, we assume that each user honestly applies an LDP algorithm to her neighbor list and sends the noisy neighbor lists to the data collector. 
However, recent studies \cite{Cao_USENIX21,Cheu_SP21} show that LDP algorithms are vulnerable to \textit{data poisoning attacks}, which inject malicious user accounts and send fake data from these accounts to degrade the accuracy of statistical analysis results or machine learning models. 
Therefore, we finally evaluate the robustness of our DPRR against data poisoning attacks. 
Specifically, we introduce data poisoning attacks to our DPRR and a defense against the attacks. 
We evaluate them through experiments and discuss the results. 
}

\smallskip
\noindent{\textbf{Our Contributions.}}~~Our contributions are as follows:
\begin{itemize}
    \item We propose the DPRR 
    for GNNs under LDP for edges. 
    Technically, we use edge sampling 
    \cite{Bera_KDD20,Eden_FOCS15,Imola_USENIX22,Wu_TKDE16} 
    after Warner's RR. 
    In particular, 
    our main technical novelty lies in what we call \textit{strategic} edge sampling, where 
    each user's sampling probability 
    is automatically tuned using the Laplacian mechanism 
    to preserve the degree information under edge LDP 
    (see Section~\ref{sub:algorithm_details}). 
    \item As explained above, 
    \colorB{the DPRR divides the privacy budget $\epsilon$ into two budgets, $\epsilon_1$ and $\epsilon_2$ ($\epsilon = \epsilon_1 + \epsilon_2$). 
    $\epsilon_1$ is for the Laplacian mechanism, whereas $\epsilon_2$ is for Warner's RR.}
    We also propose a privacy budget allocation method to make the noise in both of the mechanisms small and thereby provide high accuracy in GNNs (see Section~\ref{sub:allocation}). 
    \item We prove that our DPRR approximately preserves the degree information under edge LDP. 
    We also show that our DPRR has much smaller time and space complexity than Warner's RR (see Sections~\ref{sub:properties} and \ref{sub:time_space}). 
    \item We focus on graph classification and evaluate our DPRR using 
    three 
    social graph datasets. 
    We show that our DPRR outperforms 
    three 
    private baselines 
    (RR, \LocalLap{}, and \NPPartial{}) 
    in terms of accuracy and the RR in terms of efficiency. 
    We also compare the DPRR with a fully non-private algorithm that does not add any noise for all users, including private users (denoted by \NPFull{}). 
    For all datasets, we show that our DPRR provides accuracy close to \NPFull{} (e.g., the difference in the classification accuracy or AUC is smaller than $0.1$ or so) 
    with a reasonable privacy budget, e.g., $\epsilon = 1$ (see Section~\ref{sec:exp}). 
    \item \colorB{We finally introduce data poisoning attacks to our DPRR and a defense against the attacks. 
    We evaluate them using the three datasets and discuss the results (see Section~\ref{sec:data_poisoning_defense}).}  
\end{itemize}
Our code is available on GitHub \arxiv{\cite{DPRR-GNN}}\conference{\cite{DPRR-GNN}}. 

\smallskip
\noindent{\textbf{Technical Novelty.}}~~As described above, the technical novelty of our DPRR is two-fold: (i) strategic edge sampling that tunes each user's sampling probability to preserve the degree information under edge LDP and (ii) privacy budget allocation. 

Regarding the first point, Imola \textit{et al.} \cite{Imola_USENIX22} propose the ARR (Asymmetric RR) \cite{Imola_USENIX22}, which uses edge sampling after Warner's RR. 
Edge sampling has been widely studied to improve the scalability in 
counting triangles in a graph \cite{Bera_KDD20,Eden_FOCS15,Wu_TKDE16}. 
The authors in \cite{Imola_USENIX22} use edge sampling to improve the communication efficiency in triangle counting under LDP. 
However, they use a sampling probability common to all users and manually set the sampling probability. 
In this case, the accuracy is not improved by the sampling, because the graph structure remains destroyed. 
In statistics, random sampling is used to improve efficiency 
at the expense of accuracy. 
In fact, edge sampling in \cite{Imola_USENIX22} decreases the accuracy of triangle counting. 
In contrast, our DPRR 
provides 
higher accuracy than 
Warner's RR because it automatically tunes each user's sampling probability to preserve the degree information. 

In other words, our DPRR is different from 
the ARR \cite{Imola_USENIX22} 
(and other existing edge sampling algorithms 
\cite{Bera_KDD20,Eden_FOCS15,Wu_TKDE16}) 
in that ours is a strategic sampling algorithm to improve \textit{both accuracy and efficiency} in GNNs by 
preserving 
the graph structure under edge LDP. 
The novelty of our strategic sampling technique is not limited to the privacy literature -- there are no sampling techniques to improve both the accuracy and efficiency, even in (non-private) triangle counting or graph machine learning, to our knowledge. 

\colorB{Salas \textit{et al.} \cite{Salas_MDAI23} propose a noise-graph mechanism that 
changes $1$ (edge) to $0$ (no edge) with probability $1-p_1$ and $0$ to $1$ with probability $1-p_0$, 
where $p_1, p_0 \in [0,1]$. 
This mechanism is equivalent to the ARR in \cite{Imola_USENIX22} when $p_1 \leq p_0$. 
More specifically, let $p, q \in [0,1]$ be the parameters of Warner's RR and edge sampling, respectively, in \cite{Imola_USENIX22}; Warner's RR flips 1/0 with probability $1-p$, and edge sampling changes $1$ to $0$ with probability $1-q$. 
Then, $p_1 = p q$ and $p_0 = 1 - (1-p)q$. 
Thus, as with \cite{Imola_USENIX22}, the noise-graph mechanism in \cite{Salas_MDAI23} uses a sampling probability common to all users and manually sets the sampling probability. 
Consequently, it does not improve the accuracy of Warner's RR. 
In contrast, our DPRR automatically tunes each user's sampling probability to improve both accuracy and efficiency.}

In summary, our DPRR is new in that it adopts strategic edge sampling to improve both accuracy and efficiency. 
In addition, our privacy budget allocation method that makes the noise in the Laplacian mechanism and Warner's RR small is also new. 
We show that they are effective and outperform 
three baselines 
in terms of accuracy and efficiency.


\section{Related Work}
\label{sec:related}

\smallskip
\noindent{\textbf{DP on GNNs.}}~~In the past year or two, differentially private GNNs \cite{Jin_arXiv22,Lin_TIFS22,Mueller_arXiv22,Olatunji_arXiv21,Sajadmanesh_CCS21,Sajadmanesh_USENIX23,Wu_arXiv21_2,Wu_arXiv21,Zhang_IJCAI21} (or graph data synthesis \cite{Jian_TKDE23,Yuan_USENIX23}) have become a very active research topic. 
Most of them assume a centralized model \cite{Jian_TKDE23,Mueller_arXiv22,Olatunji_arXiv21,Sajadmanesh_USENIX23,Wu_arXiv21,Yuan_USENIX23,Zhang_IJCAI21} where the server has the entire graph (or the exact number of edges in the entire graph \cite{Wu_arXiv21}). 
They suffer from the data leakage issue and cannot be applied to decentralized (or fully decentralized) SNSs, as explained in Section~\ref{sec:intro}. 

Some studies \cite{Jin_arXiv22,Wu_arXiv21_2} focus on GNNs under LDP in different settings than ours. 
Specifically, Jin and Chen \cite{Jin_arXiv22} assume that each user has a graph and apply an LDP algorithm to a graph embedding calculated by each user. 
In contrast, we 
consider a totally different scenario, where 
each user is a node in a graph and the server does not have an edge. 
Thus, 
the algorithm in \cite{Jin_arXiv22} cannot be applied to our setting. 
Wu \textit{et al.} \cite{Wu_arXiv21_2} assume that each user has a user-item graph and propose a federated GNN learning algorithm for item recommendation. 
In their work, an edge represents that a user has rated an item. 
Therefore, their algorithm cannot be applied to our setting where a node and edge represent a user and friendship, respectively. 
We also note that federated (or collaborative) learning generally requires many interactions between users and the server \cite{Kairouz_FTML21,Shokri_CCS15}. 
In contrast, our DPRR requires only \textit{one-round} interaction. 

Sajadmanesh and Gatica-Perez~\cite{Sajadmanesh_CCS21} apply LDP to each user's feature vectors in an attributed graph. 
However, they do not hide edges, which are sensitive in a social graph. 
Consequently, their algorithm cannot be applied to unattributed graphs, where GNNs provide state-of-the-art performance \cite{Xu_ICLR19}. 

\conference{To our knowledge, 
a recent work \cite{Lin_TIFS22} is the only one that 
attempts to hide 
edges 
in GNNs under LDP. 
However, the authors in \cite{Lin_TIFS22} fail to provide a better algorithm than Warner's RR in unattributed graphs. 
Specifically, they propose to denoise a noisy adjacency matrix after applying Warner's RR by minimizing the $l_1$ norm regularized least-squares of the denoised adjacency matrix. 
However, 
when we consider a binary adjacency matrix, the optimal solution is \textit{either 
a noisy adjacency matrix obtained by Warner's RR 
or a zero matrix} (i.e., trivial solution)\footnote{Their 
optimization problem is 
$\min_{\bmA} || \tbmA - \bmA ||_F^2 + \eta ||\bmA||_1$, where 
$\bmA$ is a binary adjacency matrix, 
$\tbmA$ is a noisy adjacency matrix after Warner's RR, and 
$\eta$ is a real number. 
The optimal solution $\bmA^*$ to this problem is: 
$\bmA^* = \tbmA$ if $\eta \leq 1$ and $\bmA^* = \mathbf{0}$ (zero matrix) otherwise.
} and does not improve Warner's RR.} 

We also note that LAPGRAPH \cite{Wu_arXiv21}, which assumes the centralized model, can be modified so that it works in the local model setting, as shown in this paper. 
However, the local model version of LAPGRAPH suffers from low accuracy, as it does not preserve each user's degree information. 
In Section~\ref{sec:exp}, we show that our DPRR significantly outperforms both Warner's RR and the local model version of LAPGRAPH.

\smallskip
\noindent{\textbf{LDP.}}~~LDP has been widely studied for tabular data where each row corresponds to a user. 
The main task in this setting is statistical analysis, such as distribution estimation 
\cite{Acharya_AISTATS19,Erlingsson_CCS14,Wang_USENIX17} 
and heavy hitter estimation 
\cite{Bassily_STOC15,Qin_CCS16}. 

LDP has also been used for graph applications other than GNNs, e.g., calculating subgraph counts 
\cite{Imola_USENIX21,Imola_USENIX22}, 
estimating graph metrics \cite{Ye_TKDE21}, and generating synthetic graphs \cite{Qin_CCS17}. 
Qin \textit{et al.} \cite{Qin_CCS17} propose a synthetic graph data generation technique called LDPGen, which requires two-round interaction between users and the server. 
However, two-round interaction is impractical for many practical scenarios, as it needs a lot of user effort and synchronization -- the server must wait for all users' responses in each round. 
This is prohibitively time-consuming when the number of users is large. 
Thus, we focus on algorithms based on \textit{one-round} interaction between users and the server. 
Our DPRR is a one-round algorithm and is much more practical than LDPGen \cite{Qin_CCS17}.

Qin \textit{et al.} \cite{Qin_CCS17} also propose a one-round algorithm that applies Warner's RR to each edge. 
Similarly, the studies in \cite{Imola_USENIX21,Imola_USENIX22,Ye_TKDE21} use Warner's RR to calculate subgraph counts or graph metrics\footnote{The study in \cite{Ye_TKDE21} also proposes an algorithm to estimate the clustering coefficient based on Warner's RR, claiming that the clustering coefficient is useful for generating a synthetic graph based on the graph model BTER (Block Two-Level Erd\"{o}s-R\'{e}nyi) \cite{Seshadhri_PR12}. 
However, this claim is incorrect -- BTER does not use the clustering coefficient. Moreover, BTER has two parameters $\rho$ and $\eta$, which are determined by manual experimentation to fit the original graph (see Section IV in \cite{Seshadhri_PR12}). It is unclear how to automatically determine them.}. 
Since Warner's RR 
can also be applied to GNNs, we use it as a baseline. 
As described in Section~\ref{sec:intro}, Warner's RR makes a sparse graph dense and destroys the graph structure. 
Consequently, it does not provide high accuracy in GNNs, as shown in our experiments. 
The study in \cite{Nguyen_TDP16} reduces the number of 1s (edges) in Warner's RR by sampling without replacement. 
However, their proof of DP relies on the independence of each edge and is incorrect, as pointed out in~\cite{Imola_USENIX22}. 

Note that for categorical data of large domain size $k \gg 2$, the RR is outperformed by other LDP algorithms, such as OLH (Optimized Local Hashing) \cite{Wang_USENIX17}, OUE (Optimized Unary Encoding) \cite{Wang_USENIX17}, and HR (Hadamard Response) \cite{Acharya_AISTATS19}. 
However, it is proved in \cite{Acharya_ICML20} that they are not better than the RR in binary domains ($k=2$). 
Moreover, we consider a setting where both the input domain and the output range are binary (i.e., ``edge'' or ``no edge'') to make LDP mechanisms applicable to GNNs. 
In this setting (i.e., when output data are compressed to binary bits), all of the OLH, OUE, and HR are identical to Warner's RR, as they are symmetric. 
We also note that applying them to an entire neighbor list ($k=2^n$) results in prohibitively large noise, as $k$ is too large. 
Therefore, Warner's RR for each bit of the neighbor list has been widely used in graphs \cite{Imola_USENIX21,Imola_USENIX22,Qin_CCS17,Ye_TKDE21}. 
We also use Warner's RR as a building block.

\section{Preliminaries}
\label{sec:preliminaries}
In this section, we describe some preliminaries needed for this paper. 
Section~\ref{sub:notation} defines basic notation. 
Section~\ref{sub:LDP} reviews LDP on graphs. 
Section~\ref{sub:GNN} explains GNNs. 

\subsection{Basic Notation}
\label{sub:notation}
Let $\reals$, $\nnreals$, $\nats$, $\nnints$ be the sets of real numbers, non-negative real numbers, natural numbers, and non-negative integers, respectively. 
For $a \in \nats$, let $[a] = \{1, 2, \cdots, a\}$. 

Consider an unattributed social graph with $n\in\nats$ nodes (users). 
\colorB{Let $\calG$ be the set of 
possible graphs 
with a finite number of nodes}, and $G=(V,E) \in \calG$ be 
a graph 
with a set of nodes $V = \{v_1, v_2, \cdots, v_n\}$ 
and a set of edges $E \subseteq V \times V$. 
The graph $G$ can be either directed or undirected. 
In a directed graph, an edge $(v_i,v_j) \in E$ represents that user $v_i$ follows user $v_j$. 
In an undirected graph, an edge $(v_i,v_j)$ represents that $v_i$ is a friend with $v_j$.

A graph $G$ can be represented as an adjacency matrix $\bmA = (a_{i,j}) \in \{0,1\}^{n \times n}$, where $a_{i,j} = 1$ if and only if $(v_i,v_j) \in E$. 
Note that the diagonal elements are always $0$; i.e., $a_{1,1} = \cdots = a_{n,n} = 0$. 
If $G$ is an undirected graph, $\bmA$ is symmetric. 
Let $\bma_i = (a_{i,1}, a_{i,2}, \cdots, a_{i,n}) \in \{0,1\}^n$ be the $i$-th row of $\bmA$. 
$\bma_i$ is called the \textit{neighbor list} 
\cite{Qin_CCS17} 
of user $v_i$. 
Let $d_i \in \nnints$ be the degree 
of user $v_i$. 
Note that $d_i = ||\bma_i||_1$, i.e., the number of 1s in $\bma_i$. 
Let $\calN(v_i)$ be the set of nodes adjacent to $v_i$, i.e., $v_j \in \calN(v_i)$ if and only if $(v_i,v_j) \in E$. 

We focus on a local privacy model \cite{Imola_USENIX22,Imola_USENIX21,Ye_TKDE21,Qin_CCS17}, 
where user $v_i$ obfuscates her neighbor list $\bma_i$ using a \textit{local randomizer} $\calR_i$ and sends the obfuscated data $\calR_i(\bma_i)$ to a server. 
Table~\ref{tab:notation} shows the basic notation used in this paper. 

\begin{table}[t]
\caption{Basic notation in this paper.}
\centering
\hbox to\hsize{\hfil
\begin{tabular}{l|l}
\hline
Symbol		&	Description\\
\hline
$n$         &   Number of users.\\
$\calG$     &   Set of possible graphs.\\
$G=(V,E)$   &   Graph with users $V$ and edges $E$.\\
$v_i$       &   $i$-th user.\\
$\bmA=(a_{i,j})$	&   Adjacency matrix.\\
$\bma_i$	&		Neighbor list of user $v_i$.\\
$d_i$       &   Degree of user $v_i$.\\
$\calN(v_i)$    &   Set of nodes adjacent to user $v_i$.\\
$\calR_i$   &       Local randomizer of user $v_i$.\\
\hline
\end{tabular}
\hfil}
\label{tab:notation}
\end{table}

\subsection{Local Differential Privacy on Graphs}
\label{sub:LDP}
\noindent{\textbf{Edge LDP.}}~~The local randomizer $\calR_i$ of user $v_i$ adds some noise to her neighbor list $\bma_i$ to hide her edges. 
Here, we assume that the server and other users can be honest-but-curious adversaries and  
can obtain all edges in $G$ other than edges of $v_i$ as background knowledge. 
To strongly protect edges of $v_i$ from these adversaries, we use DP as a privacy metric. 
In graphs, there are two types of DP: 
\textit{node DP} and \textit{edge DP} \cite{Raskhodnikova_Encyclopedia16}. 
Node DP hides one node along with its edges from the adversary. 
However, many applications in the local privacy model require a user to send her user ID, and we cannot use node DP for these applications. 
Therefore, we use edge DP 
in the same way as the previous work on graph LDP \cite{Imola_USENIX21,Imola_USENIX22,Qin_CCS17,Ye_TKDE21}. 

Edge DP hides one edge between any two users from the adversary. 
Its local privacy model version, called 
\textit{edge LDP} 
\cite{Qin_CCS17}, 
is defined as follows: 
\begin{definition}[$\epsilon$-edge LDP \cite{Qin_CCS17}]\label{def:edge_LDP}
Let $\epsilon \in \nnreals$ and $i \in [n]$. 
A local randomizer $\calR_i$ of user $v_i$ with domain $\{0,1\}^n$ provides \emph{$\epsilon$-edge LDP} if and only if for any two neighbor lists $\bma_i, \bma'_i \in \{0,1\}^n$ that differ in one bit and any $s \in \mathrm{Range}(\calR_i)$, 
\begin{align}
\Pr[\calR_i(\bma_i) = s] \leq e^\epsilon \Pr[\calR_i(\bma'_i) = s].
\label{eq:edge_LDP}
\end{align}
\end{definition}
For example, the 
randomized neighbor list in \cite{Qin_CCS17} 
applies Warner's RR (Randomized Response) \cite{Warner_JASA65}, which flips 0/1 with probability $\frac{1}{e^\epsilon + 1}$, to each bit of $\bma_i$ (except for $a_{i,i}$). 
By (\ref{eq:edge_LDP}), this randomizer provides $\epsilon$-edge LDP. 

The parameter $\epsilon$ is called the privacy budget, and the value of $\epsilon$ is crucial in DP. 
By (\ref{eq:edge_LDP}), 
the likelihood of $\bma_i$ is almost the same as that of $\bma'_i$ when $\epsilon$ is close to $0$. 
However, they can be very different when $\epsilon$ is large. 
For example, it is well known that 
$\epsilon \leq 1$ or $2$ is acceptable for many practical scenarios, whereas 
$\epsilon \geq 5$ is unsuitable in most scenarios \cite{Hoshino_JJSDS20,DP_Li}. 
Based on this, we 
set $\epsilon \leq 2$ 
in our experiments. 

Edge LDP can be used to hide each user's neighbor list $\bma_i$. 
For example, in Facebook, user $v_i$ can change her setting so that anyone except for server administrators cannot see her neighbor list $\bma_i$. 
By using edge LDP with small $\epsilon$, we can hide $\bma_i$ even from the server administrators.

\smallskip
\noindent{\textbf{Relationship DP.}}~~In an undirected graph, each edge $(v_i,v_j)$ is shared by two users $v_i$ and $v_j$. 
Thus, both users' outputs can leak the information about $(v_i,v_j)$. 
Imola \textit{et al.} \cite{Imola_USENIX21} define \textit{relationship DP} to protect each edge during the whole process:
\begin{definition}[$\epsilon$-relationship DP \cite{Imola_USENIX21}]\label{def:relation_DP}
Let $\epsilon \in \nnreals$. 
A tuple of local randomizers $(\calR_1, \cdots, \calR_n)$ provides \emph{$\epsilon$-relationship DP} if and only if for any two undirected graphs $G, G' \in \calG$ that differ in one edge and any $(s_1, \ldots, s_n) \in \mathrm{Range}(\calR_1) \times \ldots \times \mathrm{Range}(\calR_n)$, 
\begin{align}
  &\Pr[(\calR_1(\bma_1), \ldots, \calR_n(\bma_n)) = (s_1, \ldots, s_n)] \nonumber\\
  &\leq e^\epsilon \Pr[(\calR_1(\bma'_1), \ldots, \calR_n(\bma'_n)) = (s_1,
  \ldots, s_n)],
\label{eq:relation_DP}
\end{align}
where $\bma_i$ (resp.~$\bma_i'$) is the $i$-th row of the adjacency matrix of $G$ (resp. $G'$).
\end{definition}

\begin{proposition} [Edge LDP and relationship DP~\cite{Imola_USENIX21}]\label{prop:edge_LDP_relation_DP} 
In an undirected graph, if each of local randomizers $\calR_1, \ldots, \calR_n$ provides $\epsilon$-edge LDP, then $(\calR_1, \ldots, \calR_n)$ provides $2\epsilon$-relationship DP. 
\end{proposition}

In Proposition~\ref{prop:edge_LDP_relation_DP}, relationship DP has the doubling factor in $\epsilon$ because the presence/absence of one edge $(v_i, v_j)$ affects two bits $a_{i,j}$ and $a_{j,i}$ in neighbor lists in an undirected graph. 

Note that even if user $v_i$ hides her neighbor list $\bma_i$, her edge with her friend $v_j$ will be disclosed when $v_j$ releases $\bma_j$. 
This is inevitable in social networks based on undirected graphs. 
For example, in Facebook, user $v_i$ can change her setting so that no one can see $\bma_i$. 
However, she can also change the setting so that $\bma_i$ is public. 
Thus, if 
user $v_i$ hides $\bma_i$ and her friend $v_j$ reveals $\bma_j$, 
their edge $(v_i,v_j)$ is disclosed. 
To prevent this, $v_i$ needs to ask $v_j$ not to reveal $\bma_j$. 

In other words, relationship DP requires some trust assumptions, unlike LDP; i.e., if a user wants to keep all her edges secret, she needs to trust her friends not to reveal their neighbor lists. 
However, even if her $k \in \nats$ friends reveal their neighbor lists, only her $k$ edges will be disclosed. 
Therefore, 
the trust assumption of relationship DP is 
much weaker than that of centralized DP, where the server can leak all edges. 

\smallskip
\noindent{\textbf{Global Sensitivity.}}~~As explained above, Warner's RR is one of the simplest approaches to providing edge LDP. 
Another well-known approach is to use global sensitivity \cite{DP}:
\begin{definition}
\label{def:sensitivity}
In edge LDP, the global sensitivity $GS_f$ of a function $f: \{0,1\}^n \rightarrow \reals$ is given by
\begin{align*}
    GS_f = \underset{\bma_i, \bma'_i \in \{0,1\}^n, \bma_i \sim \bma'_i}{\max} |f(\bma_i) - f(\bma'_i)|,
\end{align*}
where $\bma_i \sim \bma'_i$ represents that $\bma_i$ and $\bma'_i$ differ in one bit.
\end{definition}
For $b \in \nnreals$, let $\Lap(b)$ be the Laplacian noise with mean $0$ and scale $b$. 
Then, adding the Laplacian noise $\Lap(\frac{GS_f}{\epsilon})$ to $f$ provides $\epsilon$-edge LDP.

\subsection{Graph Neural Networks}
\label{sub:GNN}

\noindent{\textbf{Graph Classification.}}~~We focus on graph classification as a task of GNNs. 
The goal of graph classification is to predict a label of the entire graph, e.g., type of community, type of online discussion (i.e., subreddit~\cite{Reddit}), and music genre users in the graph are interested in. 

More specifically, we are given multiple graphs, some of which have a label. 
Let 
$\calG_l = \{G_1, \cdots, G_{|\calG_l|}\} \subseteq \calG$ 
be the set of labeled graphs, 
and $\calG_u = \{G_{|\calG_l|+1}, \cdots, G_{|\calG_l|+|\calG_u|}\} \subseteq \calG$ be the set of unlabeled graphs. 
Graph classification is a task that finds a mapping function $\phi$ that takes a graph $G \in \calG$ as input and outputs a label $\phi(G)$. 
We train a mapping function $\phi$ from labeled graphs $\calG_l$. 
Then, we can predict labels for unlabeled graphs using the trained function $\phi$. 

The GNN is a machine learning model useful for graph classification. 
Given a graph $G \in \calG$, the GNN calculates a feature vector $h_G$ of the entire graph $G$. 
Then it predicts a label based on $h_G$, e.g., by a softmax layer. 

\smallskip
\noindent{\textbf{Neighborhood Aggregation.}}~~Most GNNs use a neighborhood aggregation (or message passing) strategy, which 
updates a node feature of each node $v_i$ by aggregating node features of adjacent nodes $\calN(v_i)$ and combining them \cite{Gilmer_ICML17,GNN_Ma}. 
Specifically, each layer in a GNN has an aggregate function and a combine (or update) function. 
For $k \in \nnints$, let 
$h_i^{(k)}$ 
be a feature vector of $v_i$ at the $k$-th layer. 
Since we focus on an unattributed graph $G$, we create the initial feature vector $h_i^{(0)}$ from the graph structure, e.g., a one-hot encoding of the degree of $v_i$ or a constant value \cite{Errica_ICLR20,Xu_ICLR19}. 

At the $k$-th layer, we calculate $h_i^{(k)}$ as follows:
\begin{align}
    m_i^{(k)} &= \Agg^{(k)}(\{h_j^{(k-1)}: v_j \in \calN(v_i)\}) \label{eq:Agg}\\
    h_i^{(k)} &= \Com^{(k)}(h_i^{(k-1)}, m_i^{(k)}).  \label{eq:Com}
\end{align}
$\Agg^{(k)}$ is an aggregate function that takes feature vectors $h_j^{(k-1)}$ of adjacent nodes $\calN(v_i)$ as input and outputs a message $m_i^{(k)}$ for user $v_i$. 
Examples of $\Agg^{(k)}$ include 
a sum \cite{Xu_ICLR19}, mean \cite{Hamilton_NIPS17,Kipf_ICLR17}, and max \cite{Hamilton_NIPS17}. 
$\Com^{(k)}$ is a combine function that takes a feature vector $h_i^{(k-1)}$ of $v_i$ and the message $m_i^{(k)}$ as input and outputs a new feature vector $h_i^{(k)}$. 
Examples of $\Com^{(k)}$ include 
one-layer perceptrons 
\cite{Hamilton_NIPS17} and MLPs (Multi-Layer Perceptrons) \cite{Xu_ICLR19}. 

Note that (\ref{eq:Agg}) uses the set $\calN(v_i)$ of nodes adjacent to $v_i$ and therefore can leak the edge information. 
Since the algorithm in \cite{Sajadmanesh_CCS21} hides only node features, it reveals edge information in (\ref{eq:Agg}) and violates edge LDP. 
In other words, the algorithm in \cite{Sajadmanesh_CCS21} cannot be used to protect privacy in unattributed graphs.

For graph classification, the GNN outputs a feature vector $h_G$ of the entire graph. 
$h_G$ is calculated by aggregating feature vectors $h_i^{(K)}$ of the final iteration $K \in \nats$ as follows:
\begin{align}
    h_G &= \Read(\{h_i^{(K)}: v_i \in V\}). \label{eq:Read}
\end{align}
$\Read$ is a readout function, such as a sum \cite{Xu_ICLR19}, mean \cite{Xu_ICLR19}, and hierarchical graph pooling \cite{Ying_NIPS18}.

\section{Degree-Preserving Randomized Response}
\label{sec:proposal}
We propose a local randomizer that provides LDP for edges in the original graph while keeping high classification accuracy in GNNs. 
A simple way to provide LDP for edges is to use Warner's RR (Randomized Response) \cite{Warner_JASA65} to each bit of a neighbor list. 
However, it suffers from low classification accuracy because the RR makes a sparse graph dense and destroys a graph structure, 
as explained in Section~\ref{sec:intro}. 
To address this issue, we propose a local randomizer called the \textit{DPRR (Degree-Preserving Randomized Response)}, which provides edge LDP while preserving each user's degree. 

Section~\ref{sub:system} explains a system model assumed in our work. 
Section~\ref{sub:algorithm} describes the overview of our DPRR. 
Section~\ref{sub:algorithm_details} explains our DPRR in detail. 
Section~\ref{sub:allocation} proposes a privacy budget allocation method for our DPRR. 
Section~\ref{sub:properties} shows the degree preservation properties of our DPRR. 
Finally, Section~\ref{sub:time_space} shows the time and space complexity of our DPRR. 

\subsection{System Model}
\label{sub:system}

\begin{figure}[t]
  \centering
  \includegraphics[width=0.8\linewidth]{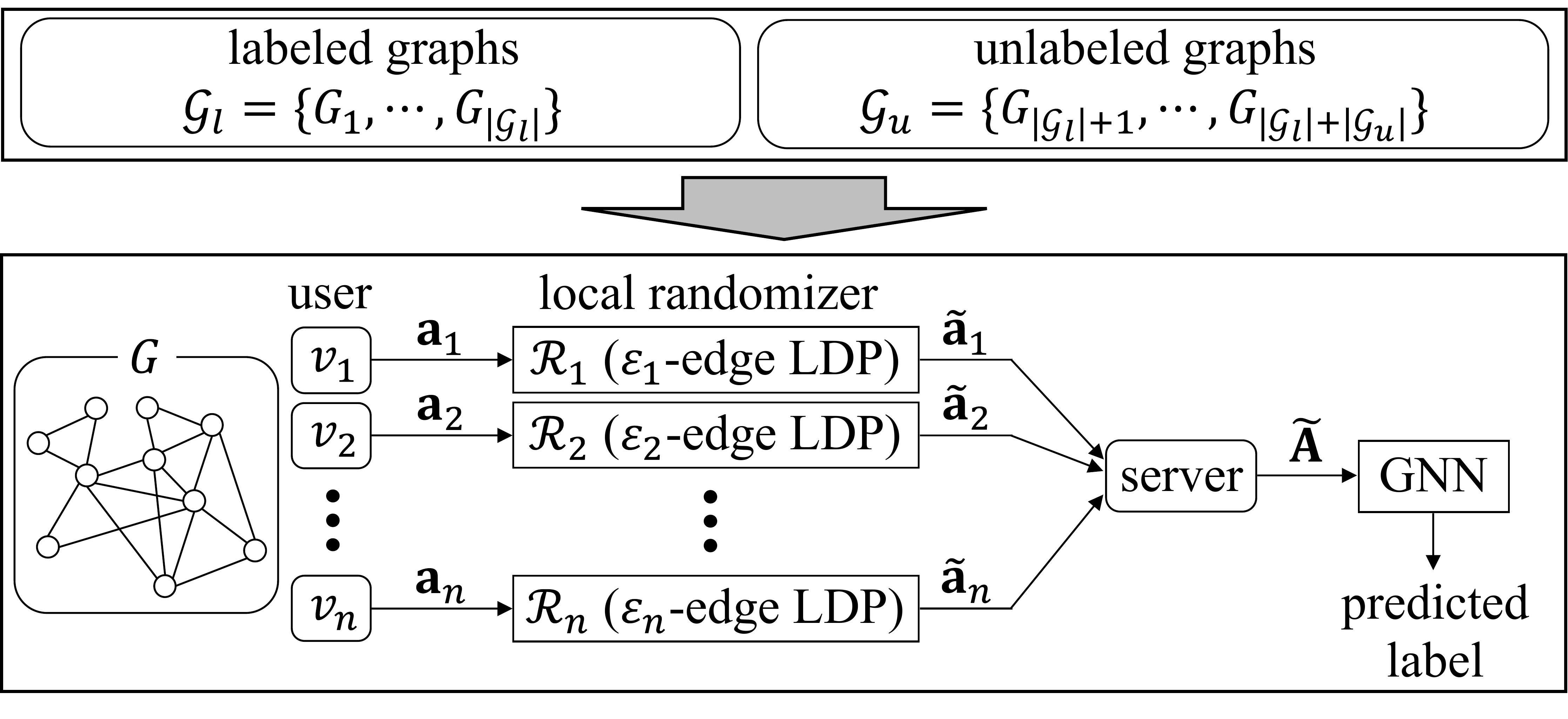}
  \vspace{-4mm}
  \caption{
  System model. For each graph $G$, users $v_1, \cdots, v_n$ send their noisy neighbor lists $\tbma_1, \cdots, \tbma_n$ providing edge LDP. 
  Then, the server 
  calculates a noisy adjacency matrix $\tbmA$ corresponding to $\tbma_1, \cdots, \tbma_n$. 
  The server trains the GNN using matrices $\tbmA$ of labeled graphs and 
  predicts a label for each unlabeled graph using its matrix $\tbmA$ and the trained GNN. 
  }
  \label{fig:system_model}
\end{figure}

Figure~\ref{fig:system_model} shows a system model 
in this paper. 
First, 
we assume that there are multiple social graphs, some of which have a label. 
We can consider several practical scenarios for this. 

For example, 
some social networks (e.g., Reddit 
\cite{Reddit}) 
provide online discussion threads. 
In this case, each discussion thread 
can be represented as 
a graph, where an edge represents that a conversation happens between two users, and a label represents a category (e.g., subreddit) of the thread. 
The REDDIT-MULTI-5K and REDDIT-BINARY \cite{Yanardag_KDD15} are real datasets in this scenario and are used in our experiments.

For another example, 
we can consider multiple 
decentralized SNSs, 
as described in Section~\ref{sec:intro}. 
Some SNSs may have a topic, such as music, game, food, and business \cite{Linkedin,Mastodon,Smule}. 
In this scenario, each SNS corresponds to a graph, and a label represents a topic. 

Based on the labeled and unlabeled social graphs, we perform graph classification privately. 
Specifically, for each graph $G = (V,E)$, each user $v_i \in V$ obfuscates her neighbor list $\bma_i \in \{0,1\}^n$ using a local randomizer $\calR_i$ providing $\epsilon_i$-edge LDP and sends her noisy neighbor lists $\tbma_i \in \{0,1\}^n$ to a server. 
Then, the server calculates a noisy adjacency matrix $\tbmA \in \{0,1\}^{n \times n}$ corresponding to $\tbma_1, \cdots, \tbma_n$. 
The server trains the GNN using noisy adjacency matrices $\tbmA$ of labeled graphs. 
Then, the server predicts a label for each unlabeled graph based on its noisy adjacency matrix $\tbmA$ and the trained GNN. 

Note that we consider a personalized setting 
\cite{Jorgensen_ICDE15}, 
where each user $v_i$ can set her privacy budget $\epsilon_i$. 
In particular, we consider the following two basic settings:
\begin{itemize}
    \item \textbf{Common Setting.} In this setting,
    all users adopt the same privacy budget $\epsilon$, 
    i.e., $\epsilon = \epsilon_1 = \epsilon_2 = \cdots = \epsilon_n$. This is a scenario assumed in most studies on DP. 
    By Proposition~\ref{prop:edge_LDP_relation_DP}, a tuple of local randomizers $(\calR_1, \ldots, \calR_n)$ provides $2\epsilon$-relationship DP in the common setting when the graph $G$ is undirected. 
    \item \textbf{Customized Setting.} In this setting, some 
    private 
    users adopt a small privacy budget (e.g., $\epsilon = 1$) and other 
    non-private 
    users make their neighbor lists public (i.e., $\epsilon = \infty$). 
    This is similar to Facebook's 
    setting. 
    Specifically, in Facebook, each user $v_i$ can change her setting so that no one (except for the server) can see $\bma_i$. 
    User $v_i$ can also change her setting so that $\bma_i$ is public. 
    Our customized setting is stronger than Facebook's setting in that a 
    private 
    user $v_i$ hides $\bma_i$ even from the server. 
\end{itemize}

Recall that 
in an undirected graph, 
even if $v_i$ hides her neighbor list $\bma_i$, her edge with $v_j$ can be disclosed when $v_j$ 
makes $\bma_j$ public. 
In other words, a tuple of local randomizers $(\calR_1, \ldots, \calR_n)$ does not provide relationship DP in the customized setting 
where one or more users are non-private. 

However, the customized setting still makes sense 
because 
it is a stronger version of Facebook's 
setting; i.e., 
our customized setting hides the neighbor list $\bma_i$ of the private user $v_i$ even from the server. 
In our customized setting, even if $k \in \nats$ friends are non-private, only $k$ edges with them will be disclosed, as described in Section~\ref{sub:LDP}.  
All edges between private users will be kept secret, even from the server.

In our experiments, we show that our DPRR is effective in both common and customized settings. 

\subsection{Algorithm Overview}
\label{sub:algorithm}

\begin{figure}[t]
  \centering
  \includegraphics[width=0.82\linewidth]{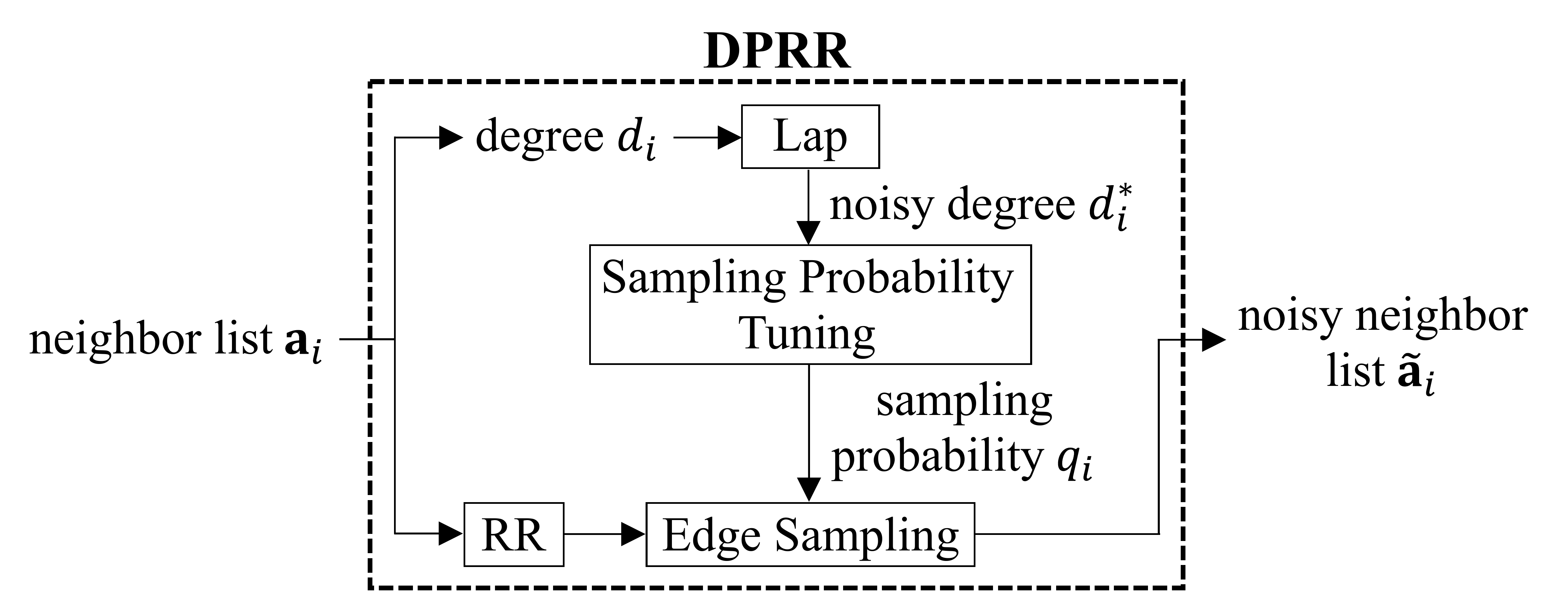}
  \vspace{-2mm}
  \caption{Overview of the DPRR (Degree-Preserving Randomized Response).}
  \label{fig:DPRR}
\vspace{2mm}
  \centering
  \includegraphics[width=0.65\linewidth]{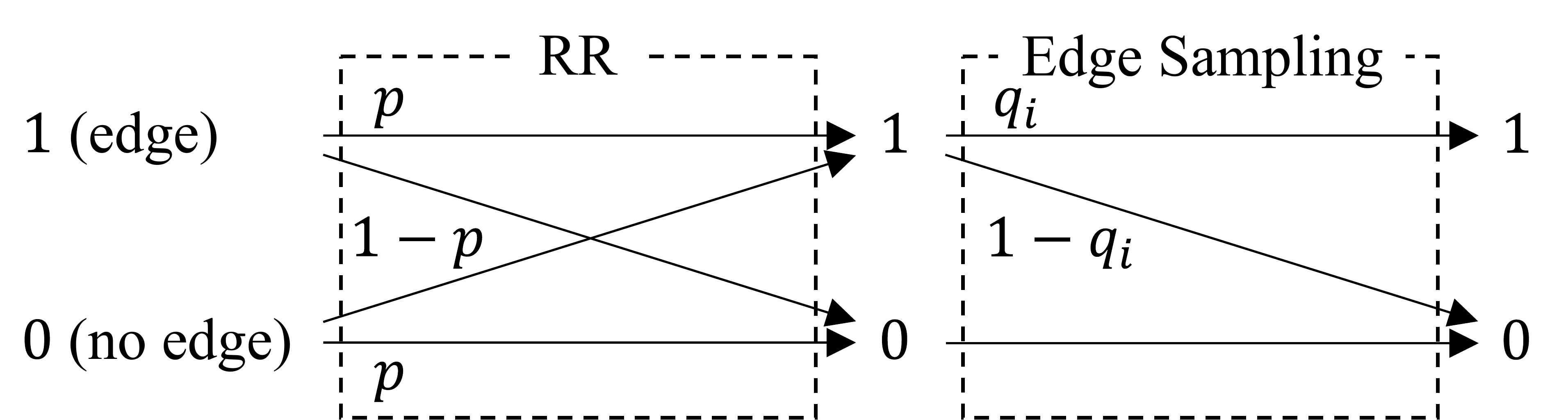}
  \vspace{-2mm}
  \caption{RR and edge sampling.}
  \label{fig:RR_edge_smpl}
\end{figure}

Figure~\ref{fig:DPRR} shows the overview of our DPRR, which takes a neighbor list $\bma_i \in \{0,1\}^n$ of user $v_i$ as input and outputs 
a noisy neighbor list $\tbma_i \in \{0,1\}^n$. 
We also show the details of the RR and edge sampling in Figure~\ref{fig:RR_edge_smpl}. 

As shown in Figures~\ref{fig:DPRR} and \ref{fig:RR_edge_smpl}, we use edge sampling 
after Warner's RR to avoid a dense noisy graph. 
For each bit of the neighbor list $\bma_i$, 
Warner's RR outputs 0/1 as is with probability $p \in [0,1]$ and flips 0/1 with probability $1-p$. 
Then, for each 1 (edge), edge sampling outputs 1 with probability $q_i \in [0,1]$ and 0 with probability $1-q_i$. 
Here, the degree information of each user is especially important in GNNs because it represents the number of adjacent nodes in the aggregate step. 
Thus, 
we carefully tune the sampling probability $q_i$ of each user $v_i$ so that \textit{the number of 1s in $\tbma_i$ is close to the original degree $d_i$} ($= ||\bma_i||_1$). 

Note that we cannot use $d_i$ itself to tune the sampling probability, because $d_i$ leaks the information about edges of $v_i$. 
To address this issue, 
we 
tune the sampling probability by replacing $d_i$ with a \textit{private estimate of $d_i$} providing edge LDP. 
\colorB{Note that the private estimate of $d_i$ is calculated locally, and therefore it provides ``local'' DP.} 
Both the RR and the private estimate of $d_i$ provide edge LDP. 
Thus, by the (general) sequential composition \cite{DP_Li}, the noisy neighbor list $\tbma_i$ is also protected with edge LDP.

Specifically, the DPRR works as follows. 
We first calculate a degree $d_i$ from the neighbor list $\bma_i$ and 
add the Laplacian noise to the degree $d_i$\footnote{We could use the Geometric mechanism, a discrete version of the Laplacian mechanism, for degrees. 
However, the Geometric mechanism does not improve the Laplacian mechanism when $\epsilon_1 < 1$ \cite{Coleman_Geo} (we set $\epsilon_1 < 0.2$ in our experiments). Thus, we use the Laplacian mechanism.}. 
Consequently, we obtain 
a private estimate 
$d_i^* \in \reals$ of $d_i$ with edge LDP. 
Then, we tune the sampling probability 
$q_i$ 
using $d_i^*$ so that the expected number of 1s in 
$\tbma_i$ is equal to $d_i$. 
Finally, we apply Warner's RR to $\bma_i$ and then randomly sample 1s with the sampling probability $q_i$. 
As a result, we obtain the noisy neighbor list $\tbma_i$ whose noisy degree $\td_i = ||\tbma_i||_1$ is almost unbiased; i.e., the expectation of $\td_i$ is almost equal to $d_i$. 
In Section~\ref{sub:algorithm_details}, we explain the details of the DPRR algorithm. 

Note that the DPRR uses two privacy budgets. 
One is for the Laplacian noise, and the other is for Warner's RR. 
In Section~\ref{sub:allocation}, we 
propose a privacy budget allocation method 
so that the variance of the noisy degree $\td_i$ is small. 
Since the noisy degree $\td_i$ is almost unbiased and has a small variance, the noisy neighbor list $\tbma_i$ preserves the degree information of user $v_i$. 
In Section~\ref{sub:properties}, we formally prove this property.

\subsection{Algorithm Details}
\label{sub:algorithm_details}
Algorithm~\ref{alg:DPRR} shows an algorithm for the DPRR. 
It assigns privacy budgets $\epsilon_1, \epsilon_2 \in \nnreals$ to the Laplacian noise and RR, respectively. 

\begin{algorithm}[t]
\caption{Degree-Preserving Randomized Response}
\begin{algorithmic}[1]
\renewcommand{\algorithmicrequire}{\textbf{Input:}}
\renewcommand{\algorithmicensure}{\textbf{Output:}}
\REQUIRE neighbor list $\bma_i \in \{0,1\}^n$, $\epsilon_1, \epsilon_2 \in \nnreals$
\ENSURE noisy neighbor list $\tbma_i \in \{0,1\}^n$
\STATE $d_i \leftarrow ||\bma_i||_1$
\STATE $d_i^* \leftarrow d_i + \Lap(\frac{1}{\epsilon_1})$
\STATE $p \leftarrow \frac{e^{\epsilon_2}}{e^{\epsilon_2}+1}$
\STATE $q_i \leftarrow \frac{d_i^*}{d_i^*(2p-1)+(n-1)(1-p)}$
\STATE $q_i \leftarrow \texttt{Proj}(q_i)$
\STATE $\tbma_i \leftarrow \texttt{RR}(\bma_i, p)$
\STATE $\tbma_i \leftarrow \texttt{EdgeSampling}(\tbma_i, q_i)$
\RETURN $\tbma_i$ 
\end{algorithmic}
\label{alg:DPRR}
\end{algorithm}

First, we add the Laplacian noise $\Lap(\frac{1}{\epsilon_1})$ to the degree $d_i$ ($= ||\bma_i||_1$) of user $v_i$ to obtain 
a private estimate 
$d_i^*$ of $d_i$: $d_i^* = d_i + \Lap(\frac{1}{\epsilon_1})$ (lines 1-2). 
Then we tune the sampling probability $q_i$ as follows:
\begin{align}
q_i = \frac{d_i^*}{d_i^*(2p-1)+(n-1)(1-p)}, 
\label{eq:q_i}
\end{align}
where $p = \frac{e^{\epsilon_2}}{e^{\epsilon_2}+1}$ (lines 3-4). 
$p$ represents the probability that Warner's RR sends an input value as is; i.e., it flips 0/1 with probability $1-p = \frac{1}{e^{\epsilon_2}+1}$. 
In Section~\ref{sub:properties}, we show that 
the noisy degree $\td_i = ||\tbma_i||_1$ becomes almost unbiased 
by setting $q_i$ 
by (\ref{eq:q_i}). 

Note that there is a small probability that $q_i$ in (\ref{eq:q_i}) is outside of $[0,1]$. 
Thus, we call the $\texttt{Proj}$ function, which projects $q_i$ onto $[0,1]$; i.e., if $q_i < 0$ (resp.~$q_i > 1$), then we set $q_i = 0$ (resp.~$q_i = 1$) (line 5). 

Next, we apply Warner's RR to a neighbor list $\bma_i$ of $v_i$. 
Specifically, we call the $\texttt{RR}$ function, which takes $\bma_i$ and $p$ as input and outputs 
a noisy neighbor list $\tbma_i$ 
by sending each bit as is with probability $p$ (line 6). 
Finally, we apply edge sampling to $\tbma_i$. 
Specifically, we call the $\texttt{EdgeSampling}$ function, which randomly samples each 1 (edge) with probability $q_i$ (line 7). 

In summary, we output $\tbma_i$ by the RR and edge sampling with the following probability: 
\begin{align}
    \colorB{\forall j \in [n]\setminus\{i\},}~ \Pr(\ta_{i,j} = 1) = \begin{cases} 
p q_i & (\text{if } a_{i,j} = 1) \\ 
(1-p) q_i & (\text{otherwise}),
\end{cases} 
\label{eq:RR_smpl}
\end{align}
where $\ta_{i,j} \in \{0,1\}$ is the $j$-th element of $\tbma_i$. 

\smallskip
\noindent{\textbf{Privacy.}}~~Below, we show the privacy property of the DPRR.
\begin{proposition}
\label{thm:DPRR_privacy}
The DPRR (Algorithm~\ref{alg:DPRR}) provides 
$\epsilon$-edge LDP, where $\epsilon = \epsilon_1 + \epsilon_2$. 
\end{proposition}

The proof is given in Appendix~\ref{sec:proof_privacy}. 
Since DP is immune to post-processing \cite{DP}, 
the GNN model trained from 
the noisy adjacency matrix $\tbmA$ provides $(\epsilon_1 + \epsilon_2)$-edge LDP as well. 

\subsection{Privacy Budget Allocation}
\label{sub:allocation}
Our DPRR uses two privacy budgets: $\epsilon_1$ for the Laplacian noise and $\epsilon_2$ for the RR. 
When $\epsilon_1$ is too small, the Laplacian noise becomes too large and causes a large variance of the noisy degree $\td_i = ||\tbma_i||_1$. 
In contrast, when $\epsilon_2$ is too small, the RR adds too much noise to each edge. 
Below, we 
propose a method 
to allocate $\epsilon_1$ and $\epsilon_2$ to avoid these two issues. 
In a nutshell, our privacy allocation method sets $\epsilon_1$ to guarantee a small variance of $\td_i$ while keeping a large value of $\epsilon_2$ to make the noise in the RR small. 

Specifically, let $n_{max} \in \nnreals$ be the maximum number of nodes (users) in all graphs 
$\calG_l$ and $\calG_u$. 
In Section~\ref{sub:properties}, we show that 
if $\epsilon_1 < \sqrt{\frac{8}{n_{max}-1}}$, the Laplacian noise is too large and causes a large variance of $\td_i$. 
Taking this into account, our privacy budget allocation method 
sets $\epsilon_1$ as follows: 
\begin{align}
    \textstyle{\epsilon_1 = \max \left\{ \sqrt{\frac{8}{n_{max}-1}}, (1-\alpha)\epsilon \right\},}
    \label{eq:epsilon_1}
\end{align}
where $\alpha$ is a constant close to $1$ ($\alpha=0.9$ in our experiments). 

This setting makes the variance of $\td_i$ small (i.e., $\epsilon_1 \geq \sqrt{\frac{8}{n_{max}-1}}$). 
It also allows the Laplacian noise to decrease with increase in $\epsilon$. 
Moreover, it makes the noise in the RR small (i.e., it makes $\epsilon_2$ large), as $\alpha$ is large. 
In our experiments, we show that this setting makes the variance of the noisy degree $\td_i$ small and provides high classification accuracy in GNNs.

\subsection{Degree Preservation Property}
\label{sub:properties}

We now show the degree preservation property of the DPRR. 
Specifically, we analyze the expectation and variance of the noisy degree $\td_i = ||\tbma_i||_1$. 

\smallskip
\noindent{\textbf{Expectation of the Noisy Degree.}}~~We 
first analyze the expectation $\mathbb{E}[\td_i]$ of the noisy degree $\td_i$ over the randomness in the Laplacian noise, the RR, and the edge sampling. 
By the law of total expectation, we have
\begin{align}
    \mathbb{E}[\td_i] = \mathbb{E}[\mathbb{E}[\td_i | d_i^*]],
    \label{eq:E_tdi}
\end{align}
where $d_i^*$ is a private estimate of $d_i$ (see line 2 of Algorithm~\ref{alg:DPRR}). 
The original neighbor list $\bma_i$ has $d_i$ 1s and $n-1-d_i$ 0s (except for $a_{i,i}$). 
By (\ref{eq:RR_smpl}), 
the DPRR 
sends 1 as is with probability $pq_i$ and flips 0 to 1 with probability $(1-p)q_i$. 
Thus, we have
\begin{align*}
    \mathbb{E}[\td_i | d_i^*] 
    &= d_i pq_i + (n - 1 - d_i) (1-p)q_i \\
    &= (d_i (2p-1) + (n-1)(1-p)) q_i \\
    &= \textstyle{\frac{d_i (2p-1) + (n-1)(1-p)}{d_i^*(2p-1)+(n-1)(1-p)} d_i^* ~~~~ \text{(by (\ref{eq:q_i}))}}. 
\end{align*}
In practice, real social graphs are sparse, which means that $d_i, d_i^* \ll n$ holds for the vast majority of nodes. 
In addition, $p = \frac{e^{\epsilon_2}}{e^{\epsilon_2}+1}$ is not close to $1$ when $\epsilon_2$ is small, e.g., $p=0.73$ when $\epsilon_2 = 1$. 
We are interested in such a value of $p$; otherwise, we cannot provide edge privacy. 
Thus, we have 
$d_i(2p-1), d_i^*(2p-1) \ll (n-1)(1-p)$, hence 
\begin{align}
    \textstyle{\mathbb{E}[\td_i | d_i^*] 
    \approx \frac{(n-1)(1-p)}{(n-1)(1-p)} d_i^* = d_i^*.}
    \label{eq:E_tdi_diast}
\end{align}
By (\ref{eq:E_tdi}) and (\ref{eq:E_tdi_diast}), we have 
\begin{align}
    \mathbb{E}[\td_i] 
    \approx \mathbb{E}[d_i^*] = d_i ~~ \text{(as the mean of $\textstyle{\Lap(\frac{1}{\epsilon_1})}$ is $0$)},
    \label{eq:E_tdi_di}
\end{align}
which means that the noisy degree $\td_i$ is almost unbiased.

\smallskip
\noindent{\textbf{Variance of the Noisy Degree.}}~~Next, we analyze the variance $\mathbb{V}[\td_i]$ of the noisy degree $\td_i$. 
By the law of total variance, we have 
\begin{align}
    &\mathbb{V}[\td_i] = \mathbb{E}\left[\mathbb{V}[\td_i | d_i^*]\right] + \mathbb{V}\left[\mathbb{E}[\td_i | d_i^*]\right].
    \label{eq:V_tdi}
\end{align}

Recall that the original neighbor list $\bma_i$ has $d_i$ 1s and $n-1-d_i$ 0s. 
By (\ref{eq:RR_smpl}), 
we have 
\begin{align}
    &\mathbb{V}[\td_i | d_i^*] \nonumber\\
    &= d_i pq_i (1 - pq_i) + (n-1-d_i) (1-p)q_i (1 - (1-p)q_i) \nonumber\\
    &\leq \textstyle{\frac{n_{max}-1}{4},}
    \label{eq:EV_tdi}
\end{align}
where the equality holds if and only if $n=n_{max}$, $p=\frac{1}{2}$ and $q_i=1$. 
By (\ref{eq:E_tdi_diast}), the second term of (\ref{eq:V_tdi}) can be written as follows: 
\begin{align}
    \textstyle{\mathbb{V}\left[\mathbb{E}[\td_i | d_i^*]\right] 
    \approx \mathbb{V}[d_i^*] 
    = \mathbb{V}[\Lap(\frac{1}{\epsilon_1})}] = \frac{2}{\epsilon_1^2}.
    \label{eq:VE_tdi}
\end{align}

By (\ref{eq:V_tdi}), (\ref{eq:EV_tdi}), and (\ref{eq:VE_tdi}), we have
\begin{align}
    \textstyle{\mathbb{V}[\td_i] 
    \approx 
    \mathbb{E}\left[\mathbb{V}[\td_i | d_i^*]\right] + \mathbb{V}\left[d_i^*\right] \leq \frac{n_{max}-1}{4} + \frac{2}{\epsilon_1^2}.}
    \label{eq:V_tdi_bound}
\end{align}
The first term of (\ref{eq:V_tdi_bound}) is caused by the randomness of the RR, 
whereas the second term of (\ref{eq:V_tdi_bound}) is caused by the randomness of the Laplacian noise. 

If $\epsilon_1 < \sqrt{\frac{8}{n_{max}-1}}$, the second term of (\ref{eq:V_tdi_bound}) is larger than the first term of (\ref{eq:V_tdi_bound}); i.e., the Laplacian noise is dominant.
Thus, our privacy allocation method sets $\epsilon_1 \geq \sqrt{\frac{8}{n_{max}-1}}$ (see (\ref{eq:epsilon_1})). 
In this case, the variance is bounded as follows: $\mathbb{V}[\td_i] \leq \frac{n_{max}-1}{2}$. 

\smallskip
\noindent{\textbf{Summary.}}~~Our DPRR makes the noisy degree $\td_i = ||\tbma_i||_1$ almost unbiased by automatically tuning the sampling probability $q_i$ by (\ref{eq:q_i}). 
In addition, our privacy allocation method makes the variance of $\td_i$ small (i.e., $\mathbb{V}[\td_i] \leq \frac{n_{max}-1}{2}$) by setting $\epsilon_1$ by (\ref{eq:epsilon_1}). 
Consequently, the noisy neighbor list $\tbma_i$ preserves the degree information of user $v_i$. 
We also show this degree-preserving property of the DPRR through experiments.

\subsection{Time and Space Complexity}
\label{sub:time_space}

\begin{table}[t]
\caption{Time and space complexity 
of three LDP algorithms per graph. 
}
\centering
\hbox to\hsize{\hfil
\begin{tabular}{c||c|c|c|c}
\hline
\multirow{2}{*}{Algorithm}	&	Time &  Time &   Space &  Space\\
        	&	(Each User) &   (Server)  &   (Each User) &  (Server)\\
\hline
DPRR        &   $O(n)$  &   $O(|E|)$    &  $O(n)$  &   $O(|E|)$\\
RR          &   $O(n)$    &   $O(n^2)$    &   $O(n)$ &   $O(n^2)$\\
\LocalLap{} &   $O(n)$  &   $O(|E|)$    &  $O(n)$    &   $O(|E|)$\\
\colorB{\NPPartial{}}    &   \colorB{$O(n)$}  &   \colorB{$O(|E|)$}    &  \colorB{$O(n)$}    &   \colorB{$O(|E|)$}\\
\hline
\end{tabular}
\hfil}
\label{tab:time_space}
\end{table}

Finally, we show that our DPRR has much smaller time and space complexity than Warner's RR. 
Table~\ref{tab:time_space} shows the time and space complexity of three LDP algorithms: the DPRR, Warner's RR applied to each bit of the adjacency matrix $\bmA$, a local model version of LAPGRAPH~\cite{Wu_arXiv21} (\LocalLap{}), \colorB{and an algorithm that uses a graph composed of only non-private users (\NPPartial{}).} 
We explain the details of \LocalLap{} in Section~\ref{sub:setup}. 
In Table~\ref{tab:time_space}, we assume that the server runs efficient GNN algorithms whose time complexity is linear in the number of edges, e.g., GIN (Graph Isomorphism Network) \cite{Xu_ICLR19}, GCN (Graph Convolutional Networks) \cite{Kipf_ICLR17}, and GraphSAGE (Graph Sample and Aggregate) \cite{Hamilton_NIPS17}. 
Some of the other GNN algorithms are less efficient; see \cite{Wu_arXiv19} for details. 

Table~\ref{tab:time_space} shows that the time and space complexity on the server side is $O(n^2)$ in Warner's RR. 
This is because the RR makes a graph dense; i.e., $|\tbmA| = O(n^2)$. 
In contrast, 
the DPRR preserves each user's degree information, and consequently, $|\tbmA| = O(|E|)$, where $|E|$ is the number of edges in the original graph $G$. 
Thus, the DPRR has the time and space complexity of $O(|E|)$ on the server side. 
Since $|E| \ll n^2$ in practice, the DPRR is much more efficient than the RR. 

For example, the Orkut social network \cite{Yang_ICDM12} includes $3072441$ nodes and $117185083$ edges. 
\colorB{The RR needs a memory size of $1$ TB to store the noisy graph on the server side. 
In contrast, the memory size of the DPRR 
required to store the noisy graph on the server side 
is only about $30$ MB, which is much smaller than that of the RR. 
This comes from the fact that the server has only sparse noisy neighbor lists generated by the DPRR.} 
In our experiments, we also show that the DPRR needs much less time for both training and classification. 

\LocalLap{} has the same time and space complexity as the DPRR. 
In our experiments, 
we show that our DPRR significantly outperforms \LocalLap{} in terms of accuracy. 

Note that each user's time and space complexity is much smaller than the server's. 
Specifically, all of the DPRR, RR, and \LocalLap{} have the time and space complexity of $O(n)$ because the length of each user's neighbor list is $O(n)$. 

\section{Experimental Evaluation}
\label{sec:exp}

Based on the theoretical properties of our DPRR in Sections~\ref{sub:properties} and \ref{sub:time_space}, we pose the following research questions:
\begin{description}[leftmargin=9.7mm]
    \item[RQ1.] 
    How does our DPRR compare with other private algorithms 
    in terms of accuracy and time complexity?
    \item[RQ2.] 
    How accurate is our DPRR compared to a non-private algorithm?
    \item[RQ3.] 
    How much does our DPRR preserve each user's degree information?
\end{description}
We conducted experiments to answer these questions. 

\subsection{Experimental Set-up}
\label{sub:setup}
\noindent{\textbf{Dataset.}}~~We 
used 
three 
unattributed social graph datasets: 

\begin{itemize}
    \item \textbf{REDDIT-MULTI-5K.}~~REDDIT-MULTI-5K \cite{Yanardag_KDD15} is a graph dataset in Reddit \cite{Reddit}, where each graph represents an online discussion thread. 
    An edge between two nodes represents that a conversation happens between the two users. 
    A label represents the type of subreddit, and there are five subreddits: worldnews, videos, AdviceAnimals, aww, and mildyinteresting. 
    \item \textbf{REDDIT-BINARY.}~~REDDIT-BINARY \cite{Yanardag_KDD15} is 
    a social graph dataset in Reddit. 
    The difference from REDDIT-MULTI-5K lies in labels. In REDDIT-BINARY, 
    a label represents a type of community, and there are two types of communities: a question/answer-based community and a discussion-based community. 
    \item \textbf{Github StarGazers.}~~Github StarGazers \cite{Rozemberczki_CIKM20} includes a social network of developers who starred popular machine learning and web development repositories until August 2019. 
    A node represents a user, and an edge represents a follower relationship. 
    A label represents the type of repository: machine learning or web. 
\end{itemize}
Table~\ref{tab:dataset} shows statistics of each graph dataset. 

\begin{table*}[t]
\caption{Statistics of graph datasets. 
``\#nodes'' represents the number of nodes in one graph. 
}
\centering
\hbox to\hsize{\hfil
\begin{tabular}{c||c|c|c|c|c|c}
\hline
\multirow{2}{*}{Dataset}    &   \#graphs    &   \#classes   &   \#nodes   &   \#nodes   &   degree   &   degree \\
    &   &   &   (max)   &   (avg)   &   (max)   &   (avg)\\
\hline
REDDIT-MULTI-5K &   4999    &   5   &   3782  &   508.5   &   2011  &   2.34\\
REDDIT-BINARY &   2000    &   2   &   3648  &   429.6  &   3062  &   2.32\\
Github StarGazers &   12725    &   2   &   957 &   113.8  &    755   &   4.12\\\hline
\end{tabular}
\hfil}
\label{tab:dataset}
\end{table*}

\smallskip
\noindent{\textbf{LDP Algorithms.}}~~For comparison, we evaluated the following four private algorithms: 
\begin{itemize}
    \item \textbf{DPRR.}~~Our proposed algorithm providing $\epsilon$-edge LDP, where $\epsilon = \epsilon_1 + \epsilon_2$ (Algorithm~\ref{alg:DPRR}). 
    As described in Section~\ref{sub:allocation}, 
    we used our privacy budget allocation method and set $\epsilon_1 = \max \left\{ \sqrt{\frac{8}{n_{max}-1}}, (1-\alpha)\epsilon \right\}$ and $\epsilon_2 = \alpha\epsilon$, where $n_{max}$ is the maximum number of nodes. \colorB{We set $\alpha=0.9$ as a default value to make the noise in the RR small (i.e., to make $\epsilon_2$ large), as described in Section~\ref{sub:allocation}. 
    We also change $\alpha$ to various values when we compare our privacy budget allocation method with a baseline that sets $\epsilon_1 = (1-\alpha)\epsilon$ and $\epsilon_2 = \alpha \epsilon$}. 
    \item \textbf{RR.}~~Warner's RR for each bit of the neighbor list \cite{Imola_USENIX21,Imola_USENIX22,Qin_CCS17,Ye_TKDE21}. 
    It flips each 0/1 with probability $\frac{1}{e^\epsilon+1}$. 
    It provides $\epsilon$-edge LDP.
    \item \textbf{\LocalLap{}.}~~A local model version of LAPGRAPH~\cite{Wu_arXiv21}. 
    LAPGRAPH is an algorithm providing edge DP in the centralized model. 
    Specifically, LAPGRAPH adds $\Lap(\frac{1}{\epsilon_1})$ to the total number of edges in a graph $G$. 
    Then it adds $\Lap(\frac{1}{\epsilon_2})$ to each element in the upper-triangular part of $\bmA$ and selects $T$ largest elements (edges), where $T$ is the noisy number of edges. 
    Finally, it outputs a noisy graph with the selected edges. 
    
    LAPGRAPH cannot be 
    used in the local model, as it needs the total number of edges in $G$ as input. 
    Thus, we modified LAPGRAPH so that each user $v_i$ sends a noisy degree $d_i^*$ ($=d_i + \Lap(\frac{1}{\epsilon_1})$) to the server 
    and the server calculates the noisy number $T$ of edges as: $T = (\sum_{i=1}^n d_i^*)/2$. 
    We call this modified algorithm \LocalLap{}. 
    \LocalLap{} provides 
    $\epsilon$-edge LDP, where $\epsilon = \epsilon_1 + \epsilon_2$. 
    We set $\epsilon_1 = \frac{\epsilon}{10}$ and $\epsilon_2 = \frac{9\epsilon}{10}$. 
    \item \textbf{\NPPartial{}.}~~An algorithm that discards neighbor lists of private users and uses only neighbor lists of non-private users. 
    Specifically, it constructs a graph composed of only non-private users and uses it as input to GNN. 
\end{itemize}
Note that \NPPartial{} is an algorithm in the customized setting and cannot be used in the common setting.

We also evaluated \NPFull{}, a fully non-private algorithm that does not add any noise for all users. 
Note that \NPFull{} does not protect privacy for private users, unlike the four private algorithms explained above. 
The accuracy of private algorithms is high if it is close to that of \NPFull{}. 

Finally, we emphasize that the algorithm in \cite{Sajadmanesh_CCS21} cannot be evaluated in our experiments, because it violates edge LDP (see Section~\ref{sub:GNN}). 

\smallskip
\noindent{\textbf{GNN Models and Configurations.}}~~Following \cite{Errica_ICLR20,Xu_ICLR19}, 
we used a constant value as the initial feature vector. 
In this case, 
GCN \cite{Kipf_ICLR17} and GraphSAGE \cite{Hamilton_NIPS17} do not perform better than random guessing, as proved in \cite{Xu_ICLR19}. 
Therefore, 
we used the GIN \cite{Xu_ICLR19}, which provides state-of-the-art performance in graph classification, as a GNN model. 
The GIN uses a sum function as 
$\Agg^{(k)}$ in (\ref{eq:Agg}), MLPs as $\Com^{(k)}$ in (\ref{eq:Com}), and a sum function as $\Read$ in (\ref{eq:Read}). 

We used the implementation in \cite{GIN_code} and used the same parameters and configurations as GIN-0 in \cite{Xu_ICLR19}, which provides the best empirical performance. 
Specifically, 
we used the mean readout as a readout function. 
We applied linear mapping and a dropout layer to a graph feature vector. 
All MLPs had two layers. 
We used the Adam optimizer. 
We applied Batch normalization to each hidden layer. 
The batch size was $64$. 

Following \cite{Errica_ICLR20,Ying_GLB21}, we tuned hyper-parameters 
via grid search. 
The hyper-parameters are: 
(i) the number of GNN layers $\in \{1,2,3,4\}$ (resp.~$\{2,3,4,5\}$) in Github StarGazers 
(resp.~REDDIT-MULTI-5K and REDDIT-BINARY); 
(ii) the number of hidden units $\in \{16,32, \allowbreak 64,128\}$; 
(iii) the initial learning rate $\in \{10^{-4}, 10^{-3}, 10^{-2}\}$; 
(iv) the dropout ratio $\in \{0, 0.5\}$.

\smallskip
\noindent{\textbf{Training, Validation, and Testing Data.}}~~For Github StarGazers, 
we followed \cite{Ying_GLB21} and randomly selected $65\%$, $15\%$, and $20\%$ of the graphs for training, validation, and testing, respectively. 
For REDDIT-MULTI-5K and REDDIT-BINARY, we randomly selected $75\%$, $10\%$, and $15\%$ for training, validation, and testing, respectively, 
as they have smaller numbers of graphs. 
We tuned the hyper-parameters explained above using the graphs for validation. 
Then we trained the GNN using the training graphs. 
Here, following \cite{Errica_ICLR20,Hassani_AAAI22}, 
we applied early stopping. 
The training was stopped at $100$ to $200$ epochs in most cases ($500$ epochs at most). 

Finally, we evaluated the classification accuracy 
and AUC (Area Under the Curve) 
using the testing graphs. 
We attempted 
ten cases for randomly dividing graphs into training, validation, and test sets 
and evaluated the average classification accuracy 
and AUC 
over the 
ten 
cases. 

\subsection{Experimental Results}
\label{sub:results}
\noindent{\textbf{Accuracy.}}~~First, we compared the accuracy of our DPRR with that of 
the three baselines (RR, \LocalLap{}, and \NPPartial{}). 
We randomly selected $\colorB{\lambda} n$ users as non-private 
users, where $\colorB{\lambda}$ is the proportion of non-private users. 
We set $\epsilon$ for private users to $\epsilon=1$ and set $\colorB{\lambda}$ to $\colorB{\lambda} = 0$, $0.1$, $0.2$, or $0.5$. 
The value $\colorB{\lambda} = 0$ corresponds to the common setting, whereas the value $\colorB{\lambda} = 0.1$, $0.2$, or $0.5$ corresponds to the customized setting. 
Then, we set $\colorB{\lambda} = 0$ (i.e., common setting) and set $\epsilon$ to $\epsilon = 0.2$, $0.5$, $1$, or $2$.

\begin{figure}[t]
    \centering
    {\small (a) REDDIT-MULTI-5K}\\
  \begin{minipage}{0.35\linewidth}
    \centering
    \includegraphics[width=\linewidth]{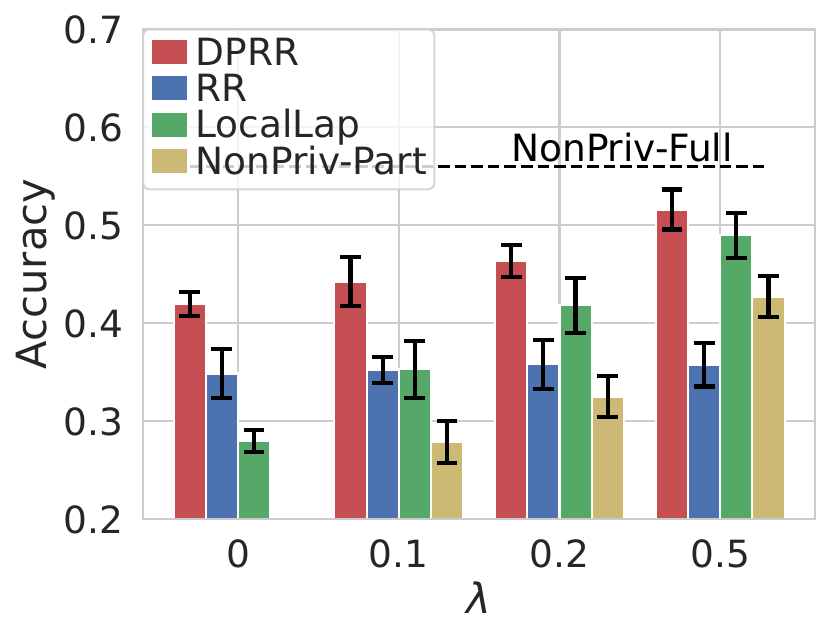}
  \end{minipage}
  \begin{minipage}{0.35\linewidth}
    \centering
    \includegraphics[width=\linewidth]{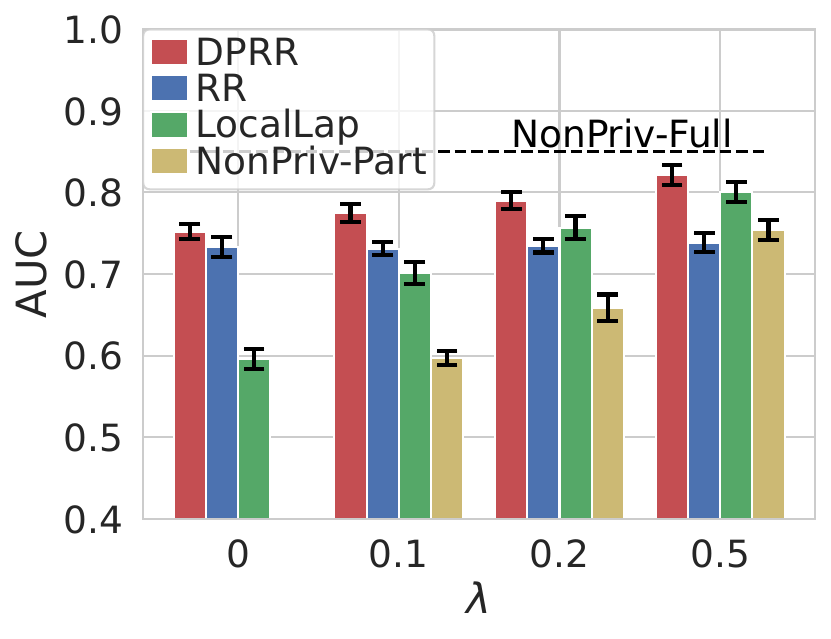}
  \end{minipage}\\
    {\small (b) REDDIT-BINARY}\\
  \begin{minipage}{0.35\linewidth}
    \centering
    \includegraphics[width=\linewidth]{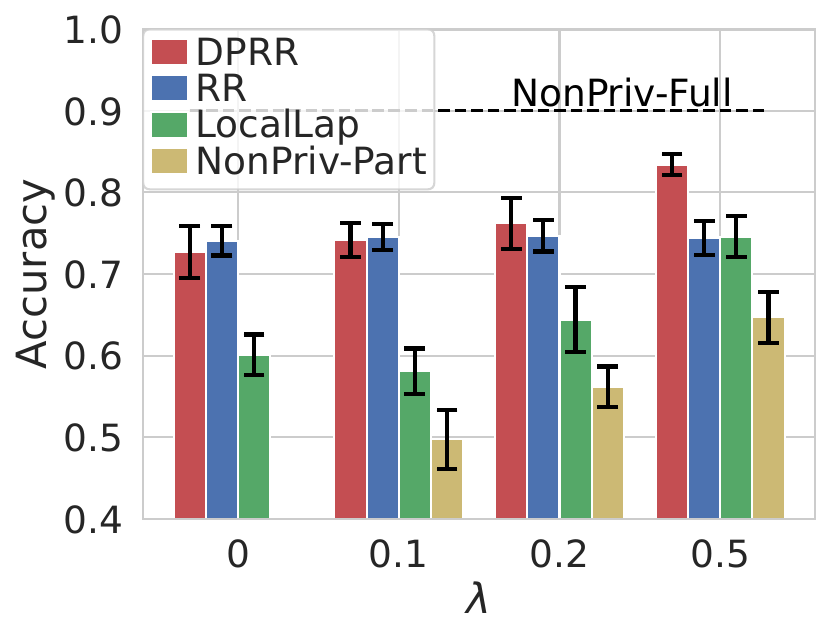}
  \end{minipage}
  \begin{minipage}{0.35\linewidth}
    \centering
    \includegraphics[width=\linewidth]{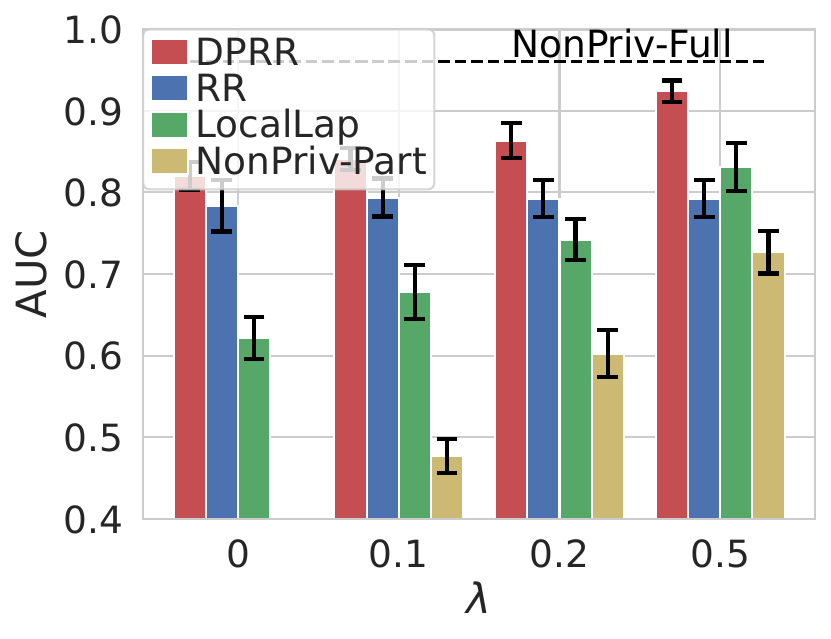}
  \end{minipage}\\
    {\small (c) Github StarGazers}\\
  \begin{minipage}{0.35\linewidth}
    \centering
    \includegraphics[width=\linewidth]{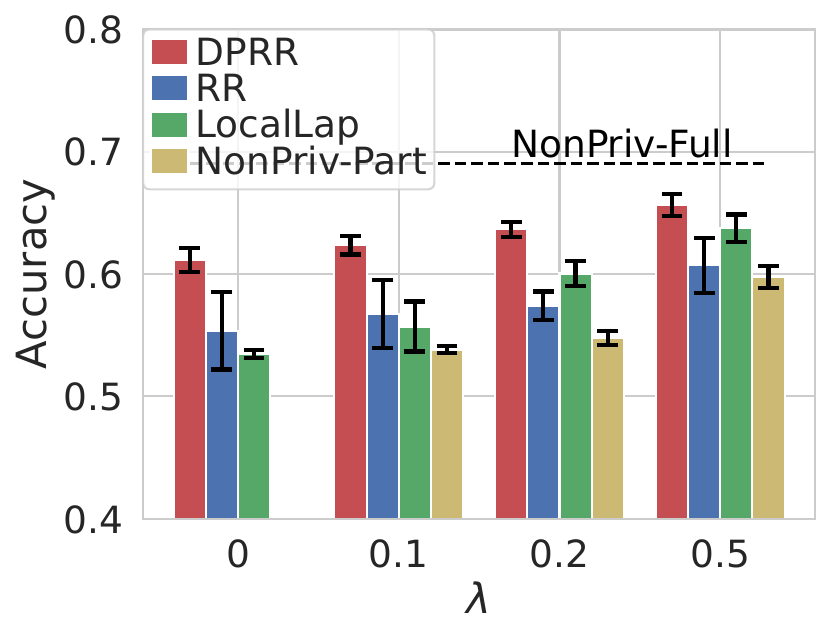}
  \end{minipage}
  \begin{minipage}{0.35\linewidth}
    \centering
    \includegraphics[width=\linewidth]{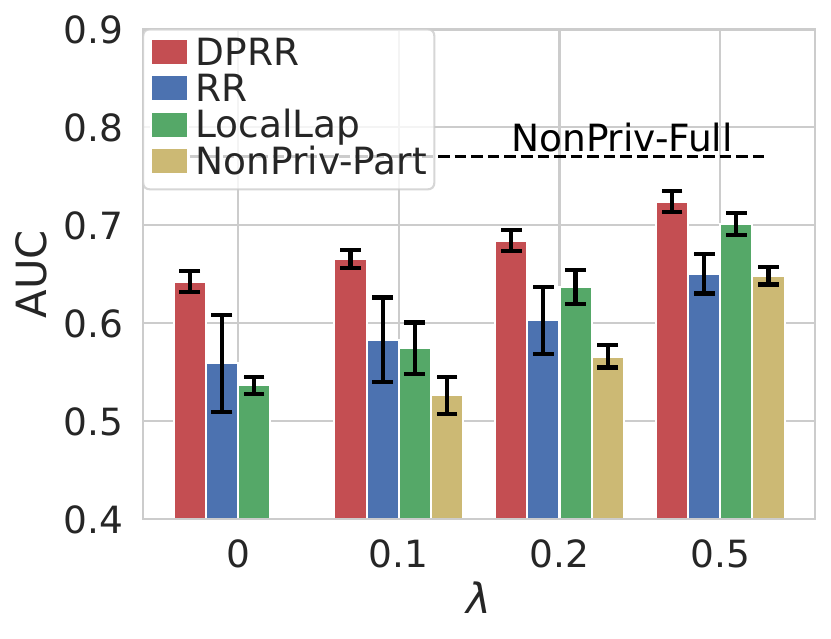}
  \end{minipage}
  \vspace{-2mm}
  \caption{Classification accuracy and AUC 
  for different \colorB{proportions $\lambda$ of non-private users} ($\epsilon=1$). 
  An error bar represents the standard deviation. 
  Note that the RR is also inefficient in terms of the training/classification time and the memory size. 
  }
  \label{fig:accuracy_alpha}
\end{figure}

\begin{figure}[t]
    \centering
    {\small (a) REDDIT-MULTI-5K}\\
  \begin{minipage}{0.35\linewidth}
    \centering
    \includegraphics[width=\linewidth]{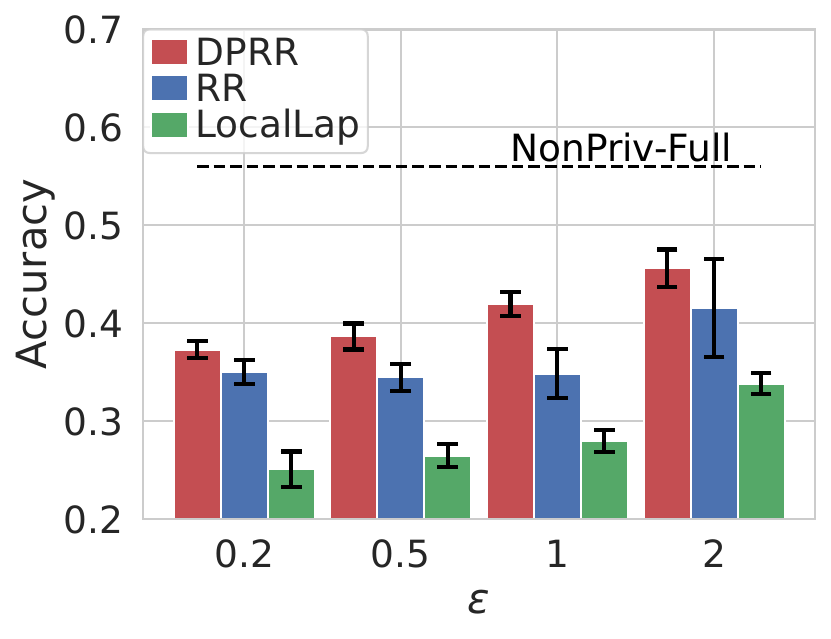}
  \end{minipage}
  \begin{minipage}{0.35\linewidth}
    \centering
    \includegraphics[width=\linewidth]{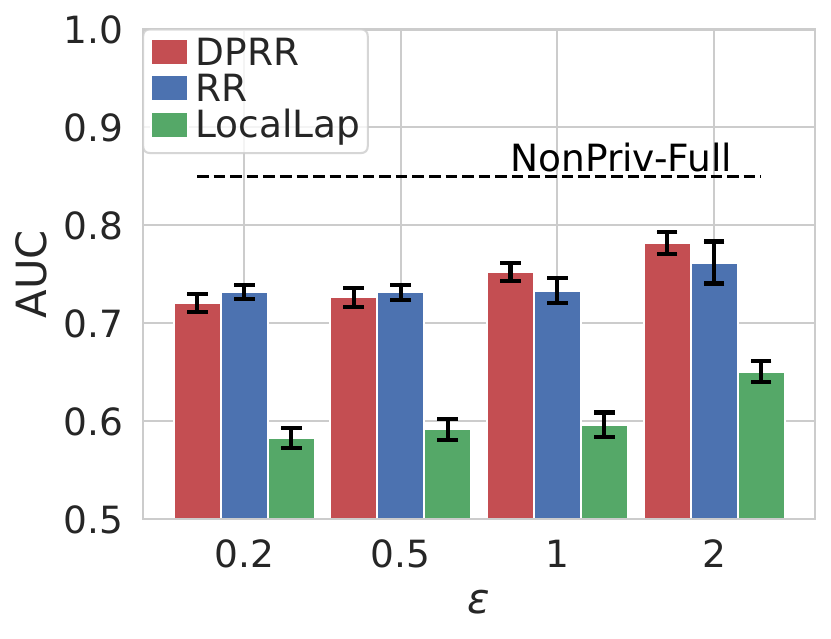}
  \end{minipage}\\
    {\small (b) REDDIT-BINARY}\\
  \begin{minipage}{0.35\linewidth}
    \centering
    \includegraphics[width=\linewidth]{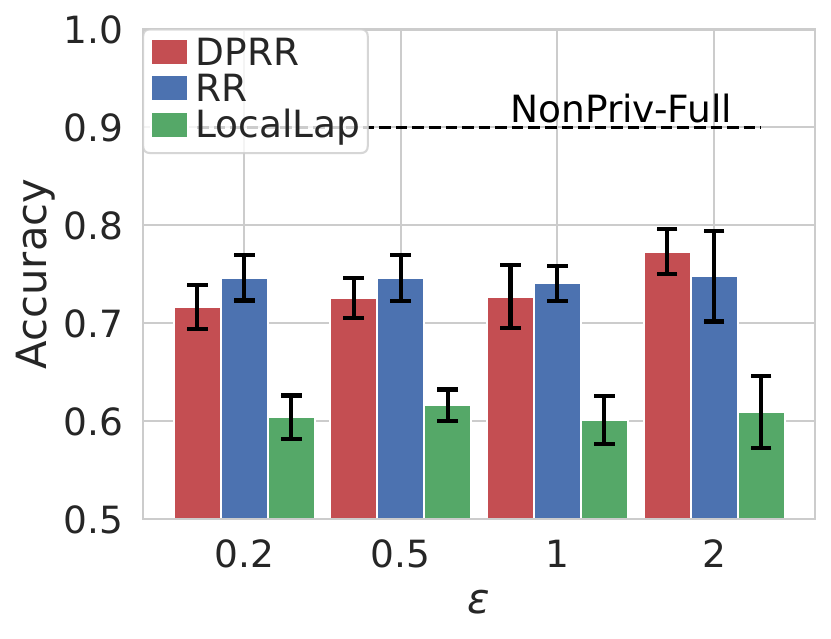}
  \end{minipage}
  \begin{minipage}{0.35\linewidth}
    \centering
    \includegraphics[width=\linewidth]{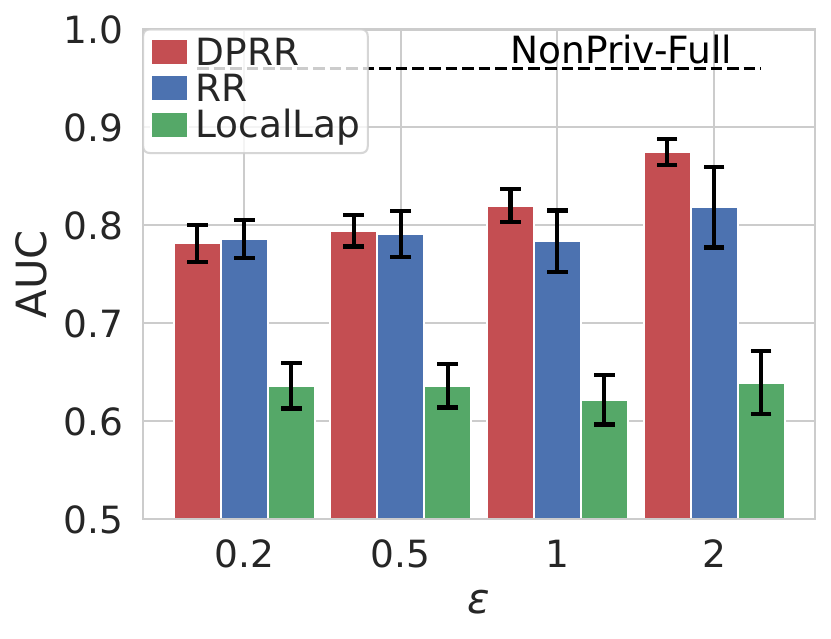}
  \end{minipage}\\
    {\small (c) Github StarGazers}\\
  \begin{minipage}{0.35\linewidth}
    \centering
    \includegraphics[width=\linewidth]{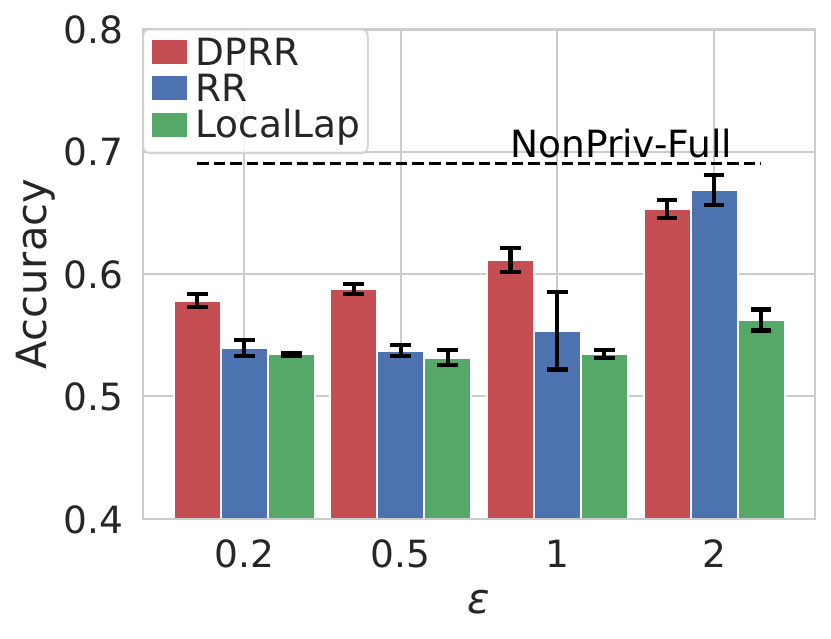}
  \end{minipage}
  \begin{minipage}{0.35\linewidth}
    \centering
    \includegraphics[width=\linewidth]{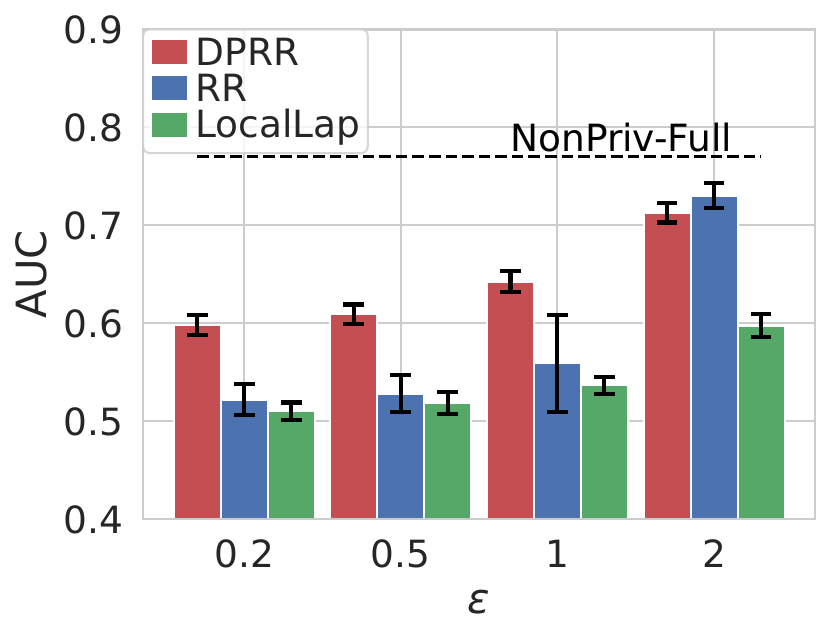}
  \end{minipage}
  \vspace{-2mm}
  \caption{Classification accuracy and AUC 
  for different privacy budgets $\epsilon$ ($\colorB{\lambda}=0$). 
  An error bar represents the standard deviation. 
  Note that the RR is also inefficient in terms of the training/classification time and the memory size. 
  }
  \label{fig:accuracy_epsilon}
\end{figure}

Figures~\ref{fig:accuracy_alpha} and \ref{fig:accuracy_epsilon} show 
the results. 
Here, a dashed line 
represents the accuracy of 
\NPFull{}. 
Note that \NPFull{} is equivalent to \NPPartial{} with $\colorB{\lambda}=1$; i.e., \NPFull{} regards all users as non-private.

\begin{figure}[t]
    \centering
    {\small (a) REDDIT-MULTI-5K (left: $\lambda=0$, right: $\lambda=0.5$)}\\
  \begin{minipage}{0.35\linewidth}
    \includegraphics[width=\linewidth]{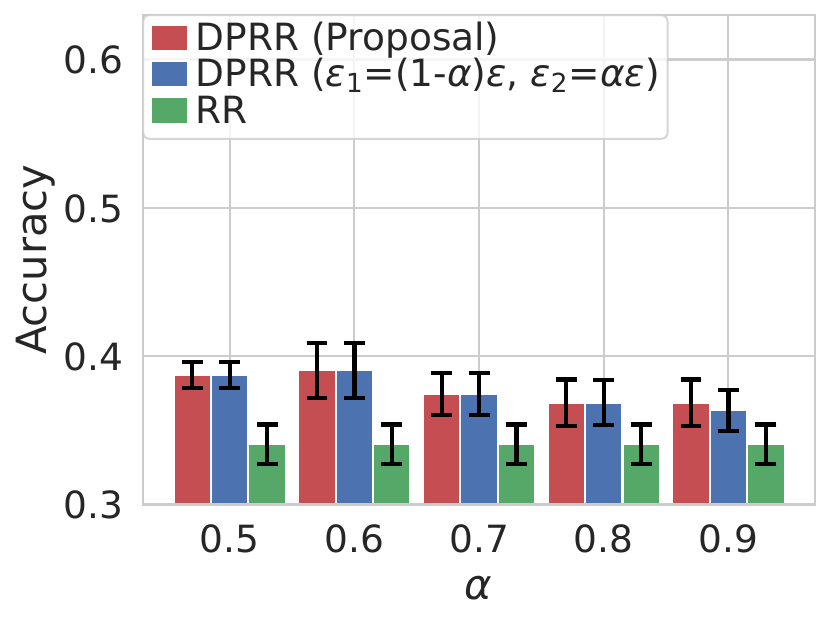}
  \end{minipage}
  \begin{minipage}{0.35\linewidth}
    \includegraphics[width=\linewidth]{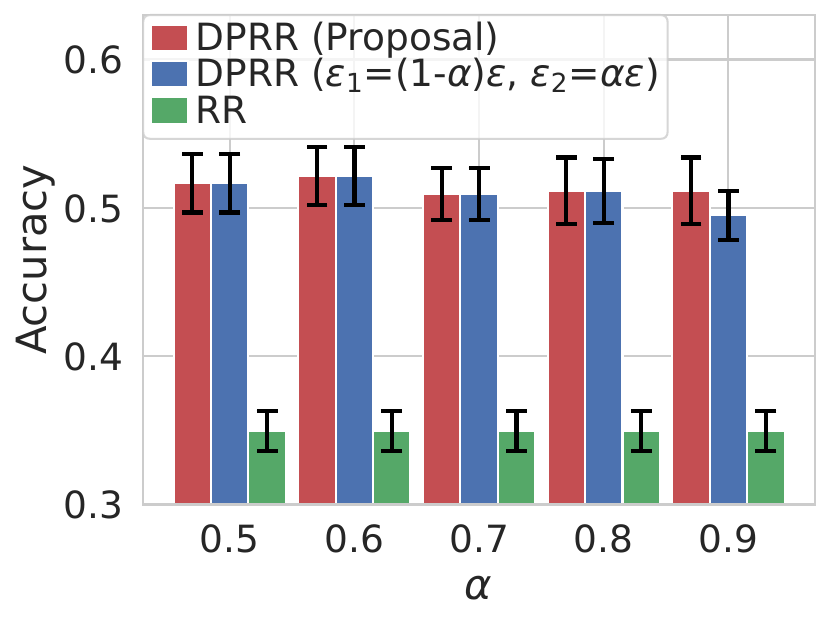}
  \end{minipage}\\
    {\small (b) REDDIT-BINARY (left: $\lambda=0$, right: $\lambda=0.5$)}\\
  \begin{minipage}{0.35\linewidth}
    \includegraphics[width=\linewidth]{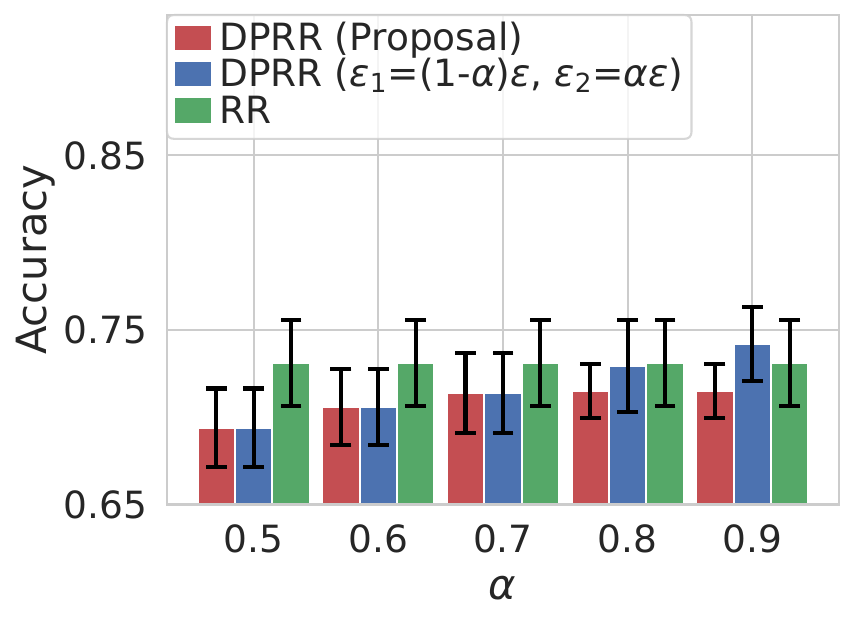}
  \end{minipage}
  \begin{minipage}{0.35\linewidth}
    \includegraphics[width=\linewidth]{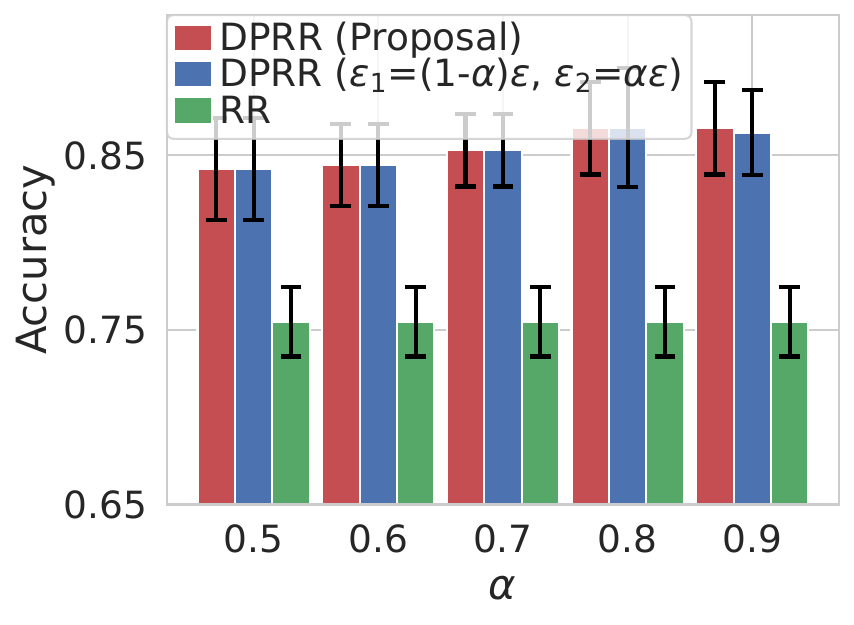}
  \end{minipage}\\
    {\small (c) Github StarGazers (left: $\lambda=0$, right: $\lambda=0.5$)}\\
  \begin{minipage}{0.35\linewidth}
    \includegraphics[width=\linewidth]{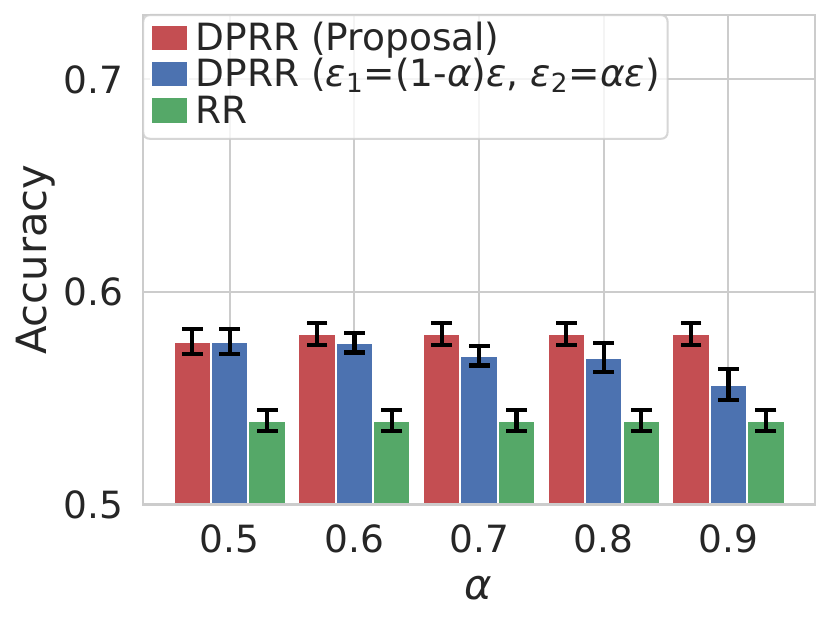}
  \end{minipage}
  \begin{minipage}{0.35\linewidth}
    \includegraphics[width=\linewidth]{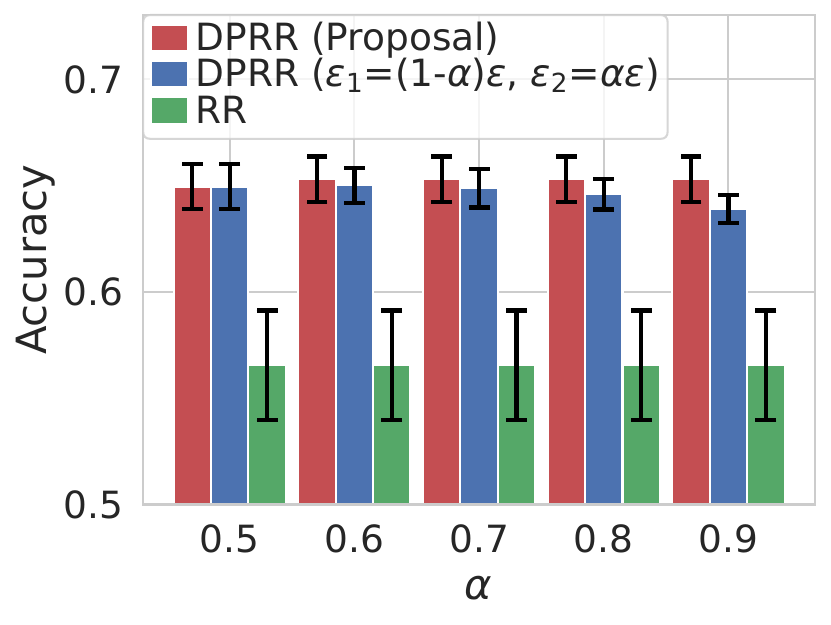}
  \end{minipage}
  \vspace{-2mm}
  \caption{\colorB{Classification accuracy for different parameters $\alpha$ in our privacy budget allocation method ($\epsilon=0.2$, $\lambda=0$ or $0.5$). An error bar represents the standard deviation.}}
  \label{fig:accuracy_budget_allocation}
\end{figure}

We observe that 
our DPRR provides the best (or almost the best) performance in all cases. 
The DPRR significantly outperforms the RR in the customized setting where $\colorB{\lambda} > 0$. 
The accuracy of the RR 
is hardly increased 
with an increase in $\colorB{\lambda}$. 
This is because the RR makes neighbor lists of private users dense. 
This ruins the neighborhood aggregation for non-private users. 
In contrast, the accuracy of our DPRR 
dramatically increases with an increase in $\colorB{\lambda}$.  
We also emphasize that the DPRR is much more efficient than the RR, as shown later.

We also observe 
that our DPRR significantly outperforms \LocalLap{} in all cases. 
One reason for this is that \LocalLap{} does not preserve each user's degree information, as shown later. 
Our DPRR also significantly outperforms \NPPartial{}, which means that the DPRR effectively uses neighbor lists of private users. 

Moreover, our DPRR provides accuracy close to \NPFull{} with a reasonable privacy budget. 
For example, 
when $\epsilon = 1$ and $\colorB{\lambda} = 0.2$, the difference in the classification accuracy or AUC is smaller than $0.1$ or so in all datasets. 
These results demonstrate the effectiveness of the DPRR. 

In our experiments, the proportion $\colorB{\lambda}$ of non-private users is the same between training graphs and testing graphs. 
In practice, $\colorB{\lambda}$ can be different between them. 
However, we note that we can make $\colorB{\lambda}$ the same between training and testing graphs by adding DP noise for some non-private users. 
For example, assume that $\colorB{\lambda} = 0.1$ in training graphs and $\colorB{\lambda} = 0.2$ in testing graphs. 
By adding DP noise for some non-private users in the testing graphs, we can make $\colorB{\lambda} = 0.1$ for both the training and testing graphs. 
Figure~\ref{fig:accuracy_alpha} shows that our DPRR is effective for all values of $\colorB{\lambda}$ including $\colorB{\lambda} = 0$.

\smallskip
\noindent{\textbf{Privacy Budget Allocation.}}~~\colorB{Next, we examined the effectiveness of our privacy budget allocation method in Section~\ref{sub:allocation}. 
Specifically, we compared our proposed method with a baseline that always sets $\epsilon_1$ and $\epsilon_2$ to $\epsilon_1 = (1 - \alpha) \epsilon$ and $\epsilon_2 = \alpha \epsilon$. 
We set $\epsilon = 0.2$ and the proportion $\lambda$ of non-private users to $\lambda = 0$ (i.e., common setting) or $0.5$ (i.e., customized setting). 
Note that we set $\epsilon$ to a small value ($=0.2$) so that a difference occurs between our proposed method and the baseline. 
Then, we changed $\alpha$ to $\alpha = 0.5$, $0.6$, $0.7$, $0.8$, or $0.9$.} 

\colorB{Figure~\ref{fig:accuracy_budget_allocation} shows the results. ``DPRR (Proposal)'' represents our DPRR with our privacy budget allocation method. 
Note that $\sqrt{\frac{8}{n_{max}-1}}$ in (\ref{eq:epsilon_1}) is $0.046$, $0.047$, and $0.091$ in REDDIT-MULTI-5K, REDDIT-BINARY, and Github StarGazers, respectively. 
Thus, ``DPRR (Proposal)'' is identical to ``DPRR ($\epsilon_1 = (1 - \alpha) \epsilon$, $\epsilon_2 = \alpha \epsilon$)'' when $\alpha \leq 0.7$, $\alpha \leq 0.7$, and $\alpha = 0.5$ in REDDIT-MULTI-5K, REDDIT-BINARY, and Github StarGazers, respectively.}

\begin{figure}[t]
    \centering
    {\small (a) REDDIT-MULTI-5K}\\
  \begin{minipage}{0.35\linewidth}
    \includegraphics[width=\linewidth]{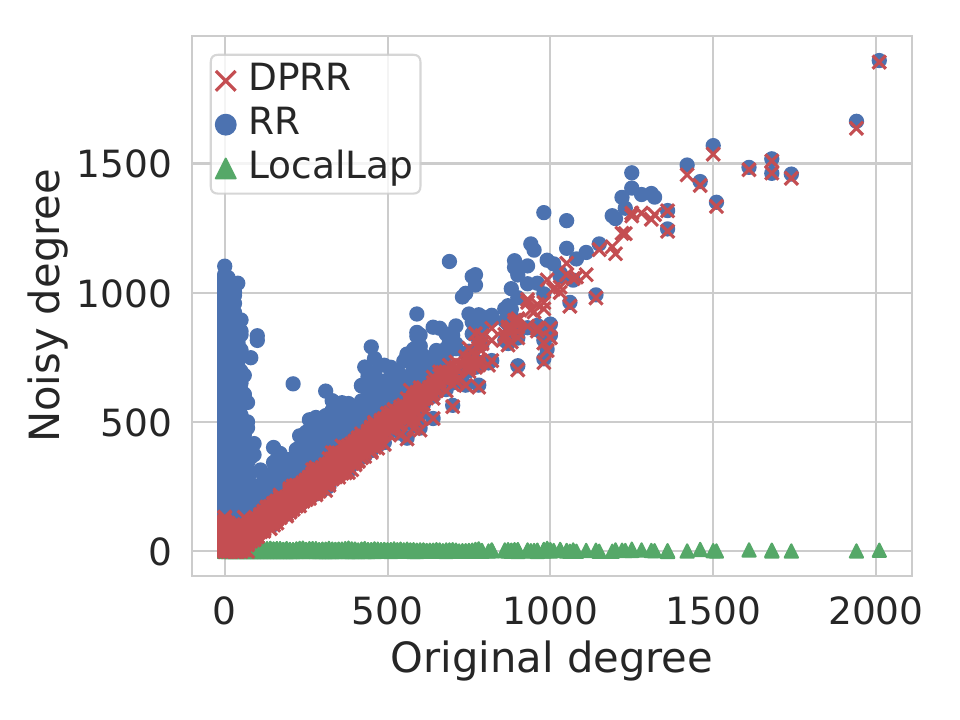}
  \end{minipage}
  \begin{minipage}{0.35\linewidth}
    \includegraphics[width=\linewidth]{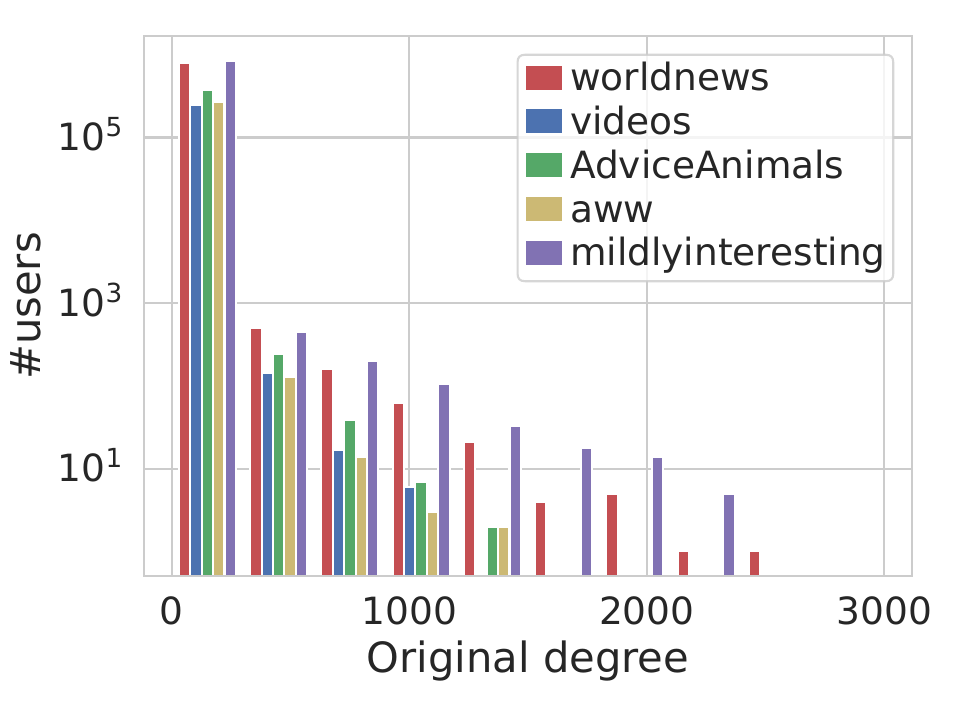}
  \end{minipage}\\
    {\small (b) REDDIT-BINARY}\\
  \begin{minipage}{0.35\linewidth}
    \includegraphics[width=\linewidth]{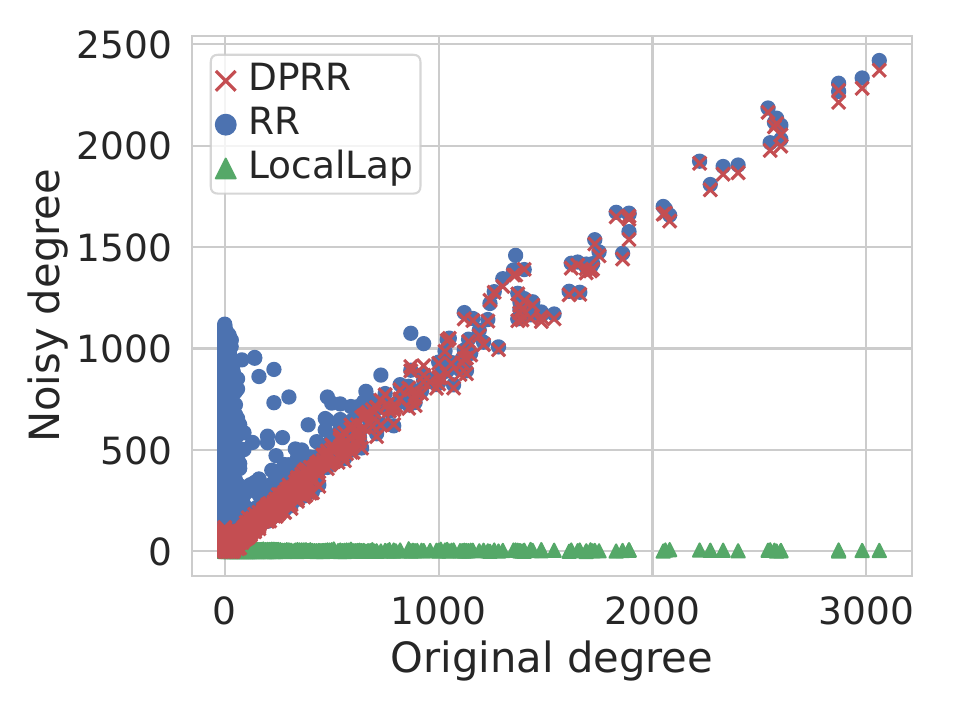}
  \end{minipage}
  \begin{minipage}{0.35\linewidth}
    \includegraphics[width=\linewidth]{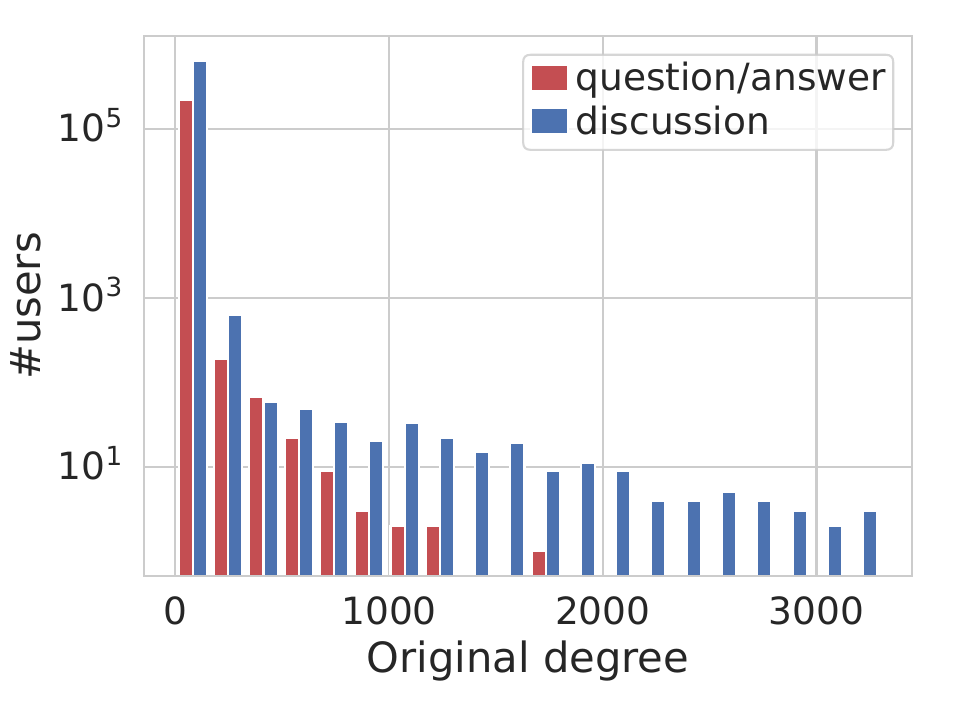}
  \end{minipage}\\
    {\small (c) Github StarGazers}\\
  \begin{minipage}{0.35\linewidth}
    \includegraphics[width=\linewidth]{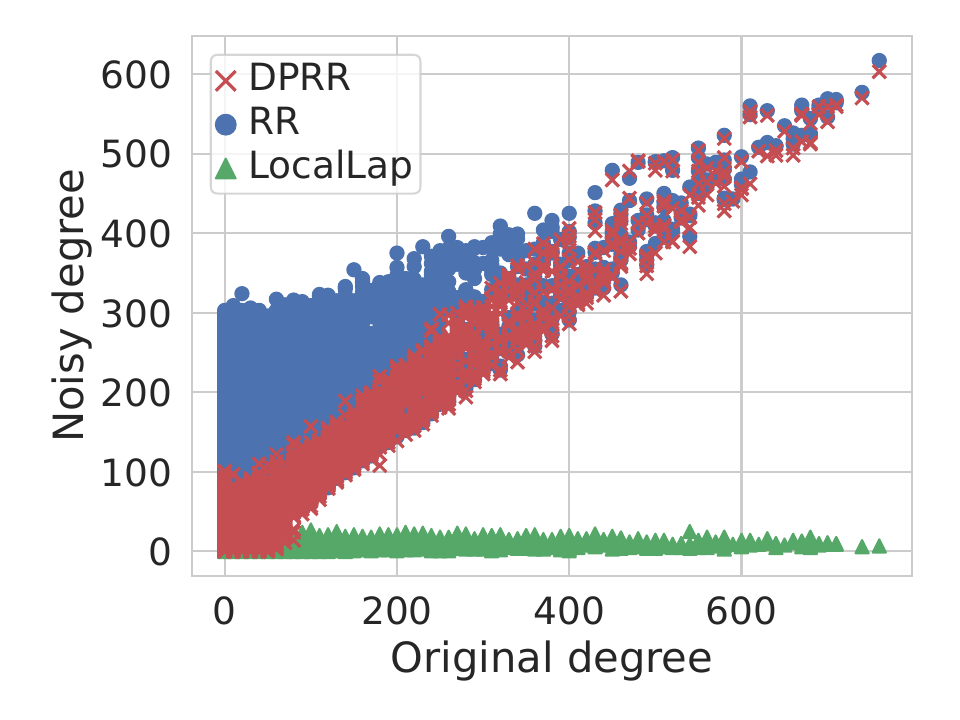}
  \end{minipage}
  \begin{minipage}{0.35\linewidth}
    \includegraphics[width=\linewidth]{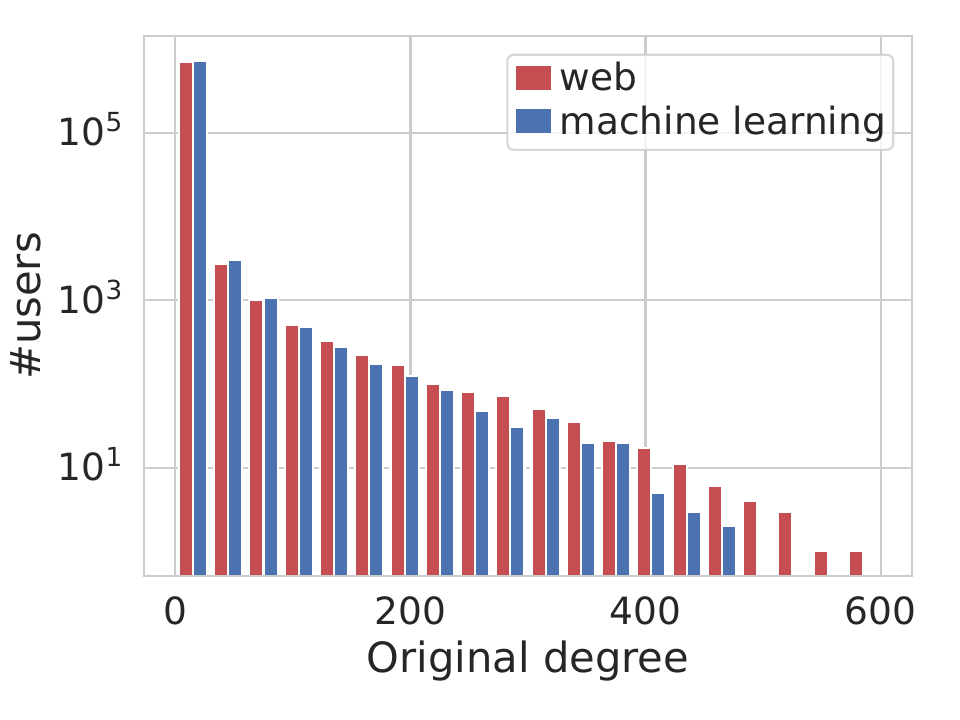}
  \end{minipage}
  \vspace{-2mm}
  \caption{Relationship between original degree $d_i$ and noisy degree $\td_i$ (left) and degree distribution for each graph type (right) ($\epsilon=1$). 
  In the left panels, each point shows the result for one user.}
  \label{fig:node_degree_scatter}
\end{figure}

\colorB{Figure~\ref{fig:accuracy_budget_allocation} (a) and (c) show that when $\alpha$ is large, our privacy budget allocation method outperforms the baseline ($\epsilon_1 = (1 - \alpha) \epsilon$, $\epsilon_2 = \alpha \epsilon$) in REDDIT-MULTI-5K and Github StarGazers. 
This is because our proposed method makes the Laplacian noise small. 
The difference is larger in Github StarGazers because the original degree $d_i$ is smaller (see Figure~\ref{fig:node_degree_scatter}) and is more susceptible to the Laplacian noise.} 

\colorB{However, the left figure of Figure~\ref{fig:accuracy_budget_allocation} (b) shows that when $\lambda = 0$, our privacy budget allocation method provides lower accuracy than the baseline ($\epsilon_1 = (1 - \alpha) \epsilon$, $\epsilon_2 = \alpha \epsilon$) in REDDIT-BINARY. 
This is because our DPRR is slightly outperformed by Warner's RR in this case. 
In other words, the degree information does not contribute much to classification accuracy. 
In this case, it might be better to set the privacy budget $\epsilon_2$ for Warner's RR as much as possible. 
In contrast, the right figure of Figure~\ref{fig:accuracy_budget_allocation} (b) shows that when $\lambda = 0.5$, our privacy budget allocation method slightly outperforms the baseline ($\epsilon_1 = (1 - \alpha) \epsilon$, $\epsilon_2 = \alpha \epsilon$) in REDDIT-BINARY. 
This is because our DPRR outperforms Warner's RR in this case.}

\colorB{In summary, our privacy budget allocation method works well, especially when the degree information contributes to classification accuracy and the original degree $d_i$ is small.}

\smallskip
\noindent{\textbf{Degree Preservation.}}~~We also examined how well the DPRR preserves each user's degree information to explain the reason that the DPRR 
outperforms the RR and \LocalLap{}. 
The left panels of Figure~\ref{fig:node_degree_scatter} show the relationship between each user $v_i$'s original degree $d_i$ and noisy degree $\td_i$. 
We also examined how the degree information of a graph correlates with its type. 
The right panels of Figure~\ref{fig:node_degree_scatter} show 
a degree distribution (i.e., distribution of $d_i$) for each graph type. 

The left panels of Figure~\ref{fig:node_degree_scatter} show 
that the DPRR preserves the original degree very well. 
This holds especially when the original degree is small, e.g., $d_i < 500$, $1000$, and $200$ in REDDIT-MULTI-5K, REDDIT-BINARY, and Github StarGazers, respectively. 
This is because the expectation of $\td_i$ is almost equal to $d_i$ (see 
(\ref{eq:E_tdi_di})) 
when $d_i\ll n$. 
In other words, the experimental results are consistent with our theoretical analysis. 
Moreover, 
the right panels of Figure~\ref{fig:node_degree_scatter} show 
that 
$d_i < 500$, $1000$, and $200$ in almost all cases in REDDIT-MULTI-5K, REDDIT-BINARY, and Github StarGazers, respectively. 
Thus, $\td_i \approx d_i$ holds for the vast majority of users in the DPRR.

The right panels of Figure~\ref{fig:node_degree_scatter} 
show that 
the degree information differs for each graph type. 
For example, 
in REDDIT-MULTI-5K, the worldnews community tends to have a larger degree than the videos community; i.e., the worldnews community tends to attract more people.

The left panels of Figure~\ref{fig:node_degree_scatter} show 
that 
the RR and \LocalLap{} do not preserve the degree information for the vast majority of users. 
The RR makes neighbor lists of these users dense and destroys the graph structure. 
\LocalLap{} makes neighbor lists sparse, irrespective of the original degrees. 
In contrast, the DPRR preserves the degree information for these users. 
This 
explains 
why 
the DPRR 
outperforms the RR and \LocalLap{}.

\smallskip
\noindent{\textbf{Training/Classification Time.}}~~Finally, we measured the time for training and classification 
on the server side 
in the DPRR, RR, and \LocalLap{}. 
Here, we used two datasets: 
REDDIT-BINARY and a synthetic graph dataset based on the BA (Barab\'{a}si-Albert) model~\cite{NetworkScience}. 
The BA model is a graph generation model that has a power-law degree distribution. 
It generates a synthetic graph by adding new nodes one by one. 
Each node has $m \in \nats$ new edges, and each edge is connected to an existing node with probability proportional to its degree. 
The average degree of the BA graph is $2m$. 
We used the NetworkX library \cite{Hagberg_SciPy08} to generate the BA graph. 

For each dataset, we evaluated the relationship between the training/classification time and the graph size. 
Specifically, in REDDIT-BINARY, we used $75\%$ of the graphs (i.e., $1500$ graphs) and $15\%$ of the graphs (i.e., $300$ graphs) for training and classification, respectively, as explained in Section~\ref{sub:setup}. 
For each graph, we randomly sampled $\gamma n$ nodes, where $\gamma \in [0,1]$ is a sampling rate, and used a subgraph composed of the sampled nodes. 
In the BA graph dataset, we generated $1000$ and $200$ graphs for training and classification, respectively. 
We set $m=3$ or $5$ and 
$n=1000$, $2000$, $3000$, or $4000$. 
We evaluated the relationship between the run time and $\gamma$ (resp.~$n$) in REDDIT-BINARY (resp.~BA graph dataset). 

We measured the run time using 
a supercomputer in \cite{ABCI}. 
We used one computing node, which consists of two Intel Xeon Platium 8360Y processors (2.4 GHz, 36 Cores) and 512 GiB main memory. 
For training, we measured the time to run $100$ epochs because the training was stopped at $100$ to $200$ epochs in most cases, as described in Section~\ref{sub:setup}. 
For classification, we measured the time to classify all graphs for classification. 
We used the implementation in \cite{GIN_code} as a code of GNN. 
Note that the DPRR, RR, and \LocalLap{} only differ in the input to GNN, and therefore the comparison is fair.

\begin{figure}[t]
  \centering
  \begin{minipage}{0.35\linewidth}
    \centering
    \includegraphics[width=\linewidth]{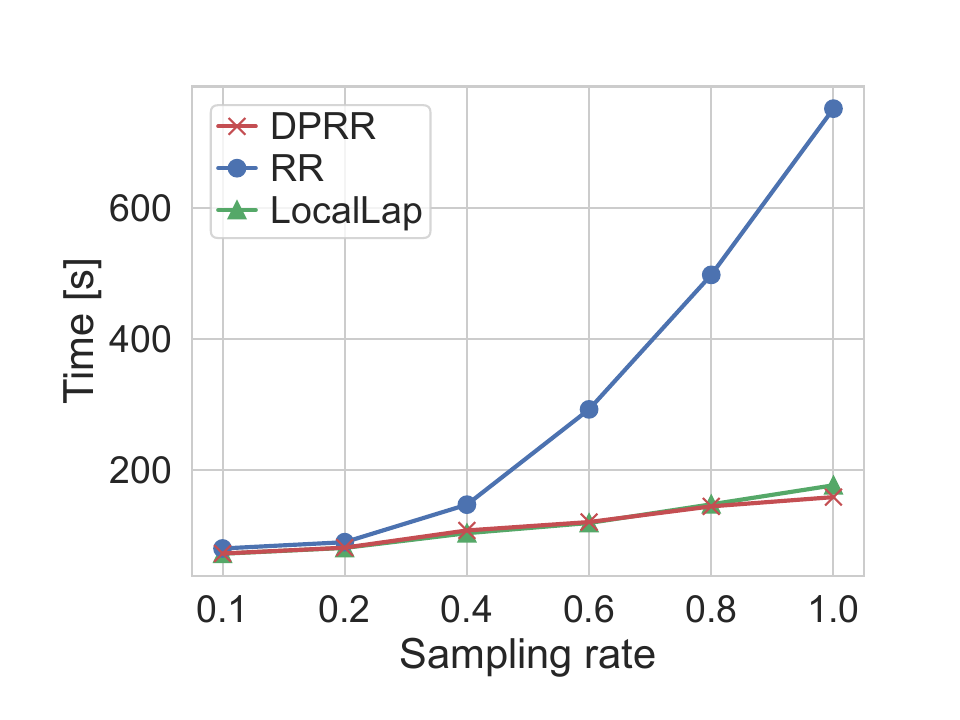}
  \end{minipage}
  \begin{minipage}{0.35\linewidth}
    \centering
    \includegraphics[width=\linewidth]{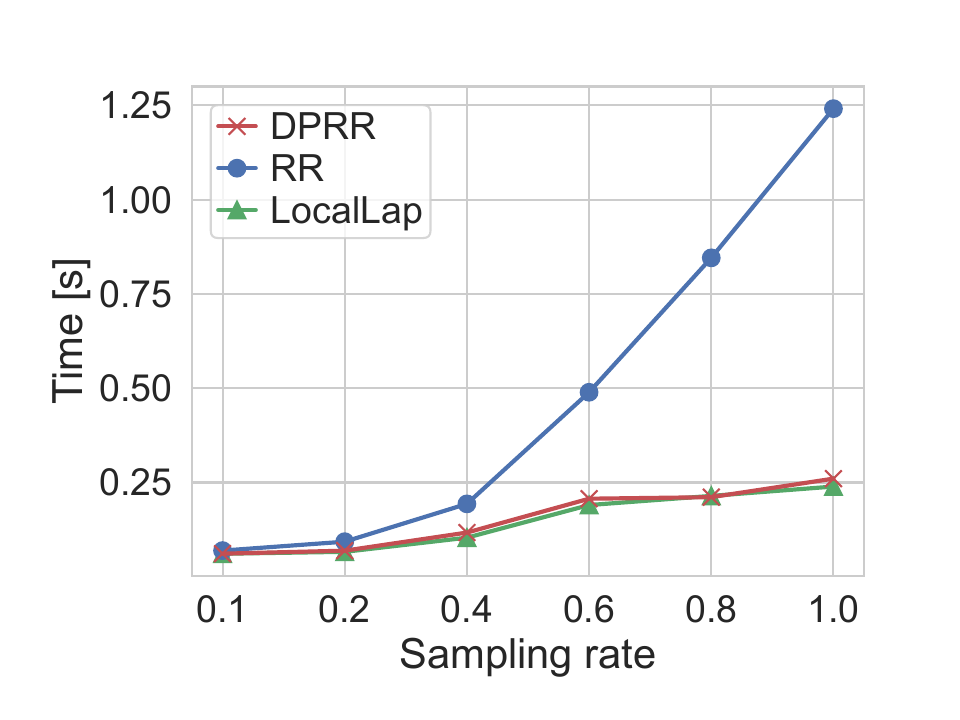}
  \end{minipage}
  \vspace{-2mm}
  \caption{Run time vs. sampling rate $\gamma$ in REDDIT-BINARY (left: training time, right: classification time).}
  \label{fig:time}
\end{figure}
\begin{figure}[t]
    \centering
    {\small (a) $m=3$}\\
  \begin{minipage}{0.35\linewidth}
    \includegraphics[width=\linewidth]{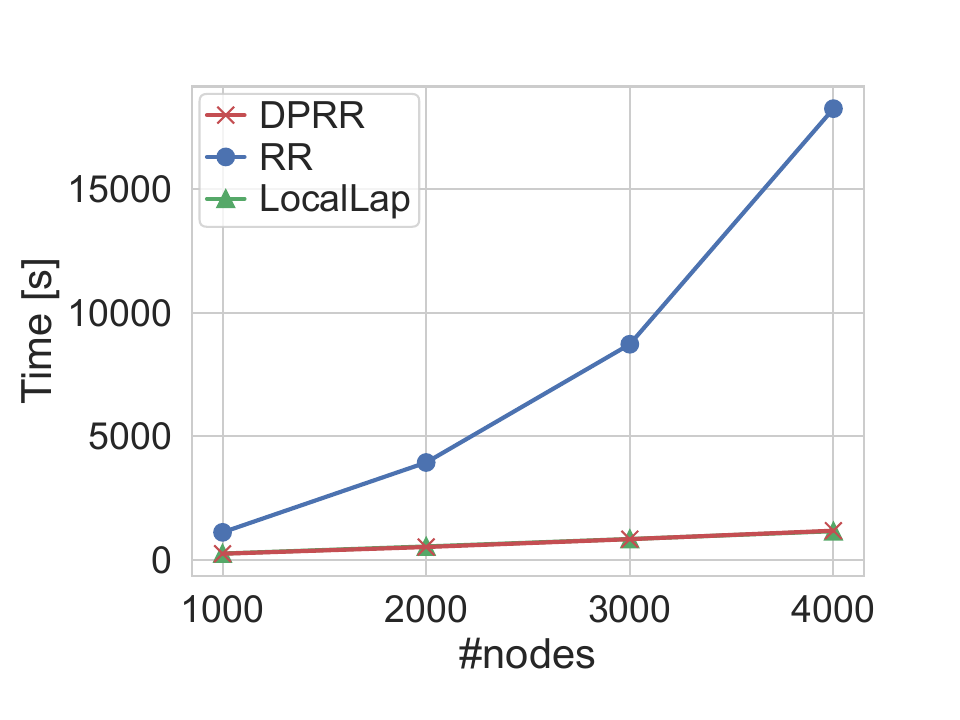}
  \end{minipage}
  \begin{minipage}{0.35\linewidth}
    \includegraphics[width=\linewidth]{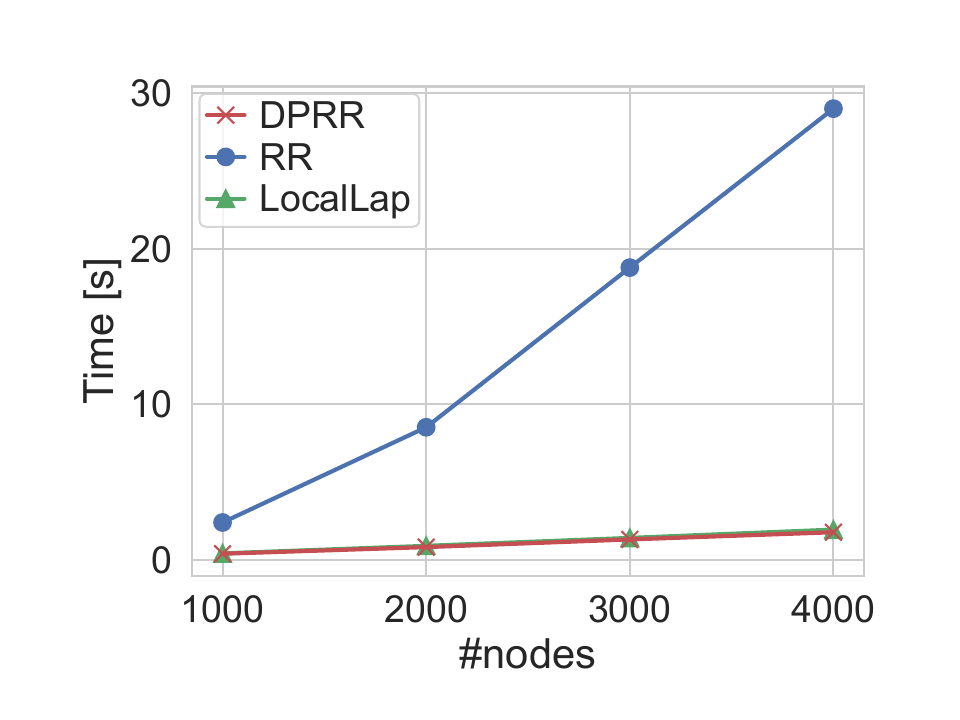}
  \end{minipage}\\
    {\small (b) $m=5$}\\
  \begin{minipage}{0.35\linewidth}
    \includegraphics[width=\linewidth]{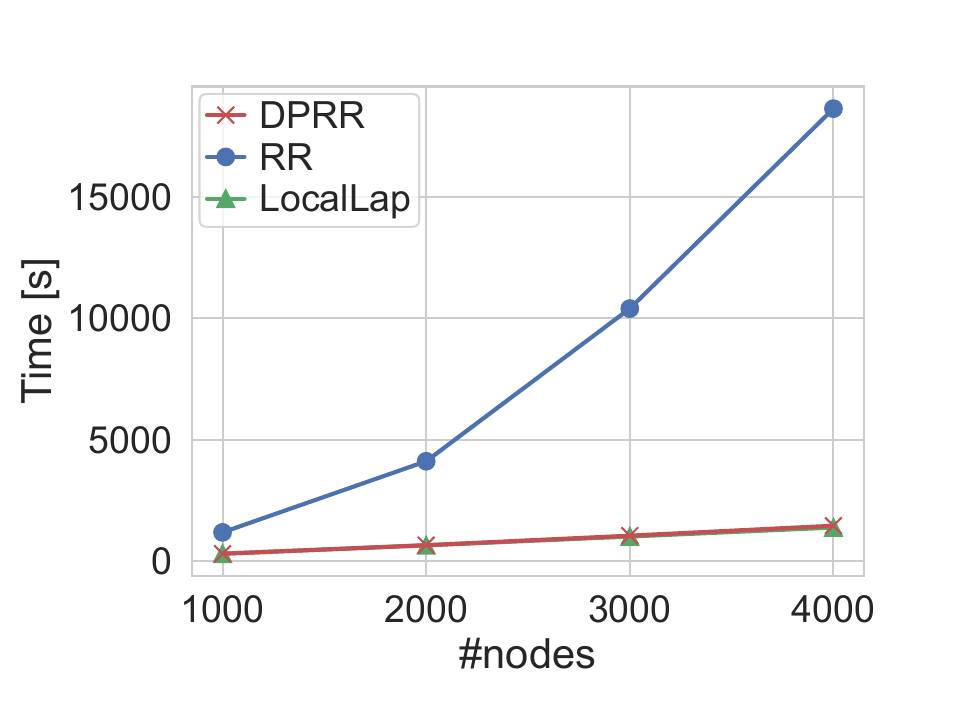}
  \end{minipage}
  \begin{minipage}{0.35\linewidth}
    \includegraphics[width=\linewidth]{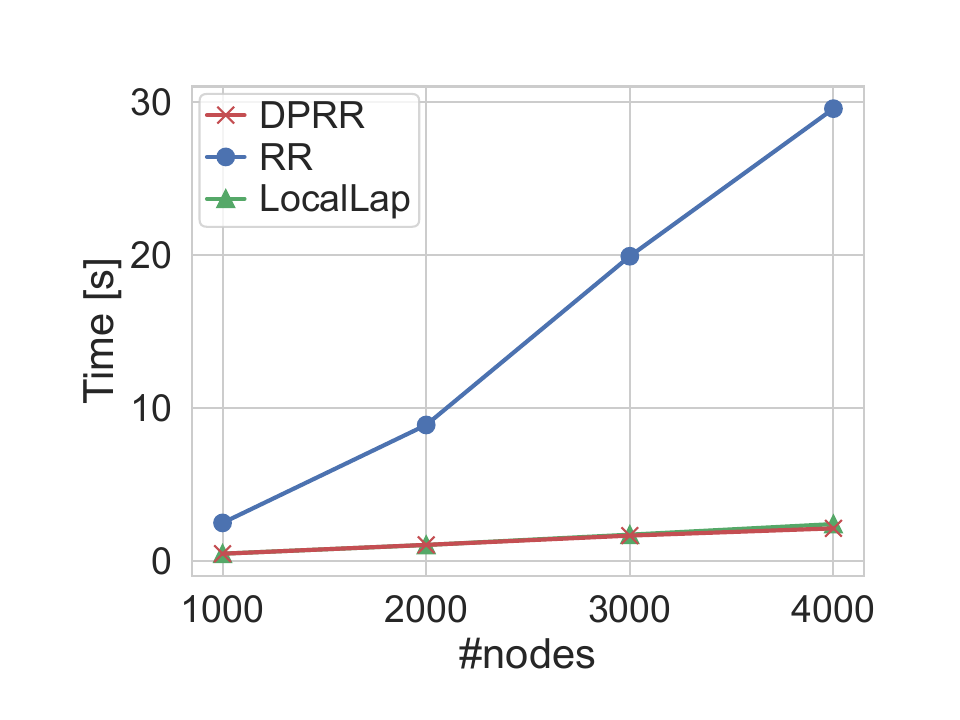}
  \end{minipage}
  \vspace{-2mm}
  \caption{Run time vs. \#nodes $n$ in the BA graphs (left: training time, right: classification time).}
  \label{fig:time2}
\end{figure}

Figures~\ref{fig:time} and \ref{fig:time2} show the results. 
The run time of the RR is large and almost quadratic in the graph size, i.e., the sampling rate $\gamma$ in Figure~\ref{fig:time} and the number $n$ of nodes in Figure~\ref{fig:time2}. 
In contrast, the run time of the DPRR and \LocalLap{} is much smaller than the RR. 
Figure~\ref{fig:time2} shows that when the average degree is constant ($=2m$), the run time of the DPRR and \LocalLap{} is almost linear in $n$. 
These results are consistent with Table~\ref{tab:time_space}. 

We can also estimate the run time for larger $n$ based on Figure~\ref{fig:time}. 
For example, Figure~\ref{fig:time} (c) shows that 
when $m=5$ and $n=4000$, the training time 
of the RR and our DPRR is about $19000$ and $1500$ seconds, respectively. 
Thus, when $m=5$ and $n=40000$, the training time of the RR and our DPRR is estimated to be about $21$ days ($=19000 \times 10^2$ seconds) and $4$ hours ($=1500 \times 10 / 3600$ seconds), respectively. 
Therefore, our DPRR is much more efficient and practical than the RR. 

\smallskip
\noindent{\textbf{Summary.}}~~In summary, our answers to the three research questions at the beginning of Section~\ref{sec:exp} are as follows. 
[RQ1]: Our DPRR is much more efficient than the RR and provides higher accuracy than the RR, especially in the customized setting. 
Our DPRR also provides much higher accuracy than the other private baselines (\LocalLap{} and \NPPartial{}) in terms of accuracy. 
[RQ2]: Our DPRR provides accuracy close to a non-private algorithm (\NPFull{}) with a reasonable privacy budget, e.g., $\epsilon=1$. 
[RQ3]: Our DPRR preserves each user's degree information very well, 
whereas the RR and \LocalLap{} do not. 
In addition, the degree information of a graph correlates with its type. 
These results explain 
why 
the DPRR outperforms the RR and \LocalLap{} in terms of accuracy. 

\section{\colorB{Data Poisoning Attacks and Defenses}}
\label{sec:data_poisoning_defense}
\colorB{As with most of the existing work on LDP (e.g., \cite{Acharya_AISTATS19,Bassily_STOC15,Erlingsson_CCS14,Imola_USENIX21,Imola_USENIX22,Qin_CCS16,Qin_CCS17,Wang_USENIX17,Ye_TKDE21}), we have so far assumed that users are honest. 
That is, we assumed that each user $v_i$ honestly applies a local randomizer $\calR_i$ to her neighbor list $\bma_i$ and reports the noisy neighbor lists $\tbma_i$. 
However, recent studies \cite{Cao_USENIX21,Cheu_SP21} show that LDP algorithms are vulnerable to data poisoning attacks, as described in Section~\ref{sec:intro}. 
Therefore, we finally evaluate the robustness of our DPRR against data poisoning attacks.} 

\colorB{Section~\ref{sub:data_poisoning} introduces 
a general data poisoning attack to our DPRR, which includes the \textit{all-ones attack} and the \textit{random attack} as concrete examples. 
Section~\ref{sub:defenses} introduces a defense against the attacks that can be applied to both directed and undirected graphs. 
Section~\ref{sub:evaluation} evaluates the all-ones attack, the random attack, and the defense through experiments and discusses the results.} 

\subsection{\colorB{Data Poisoning Attacks}}
\label{sub:data_poisoning}
\noindent{\textbf{\colorB{Threat Model.}}}~~\colorB{Assume that $\beta n$ users are malicious, where $\beta \in [0,1]$ is the proportion of malicious users. 
Each malicious user $v_i$ can change her noisy neighbor list $\tbma_i$ to an arbitrary value. 
This is referred to as the general attack \cite{Cheu_SP21}. 
The adversaries' goal is to degrade the classification accuracy of the GNN as much as possible.}

\smallskip
\noindent{\textbf{\colorB{Attack Algorithms.}}}~~\colorB{To achieve the above goal, we consider the following attack algorithm. 
Each malicious user $v_i$ applies a local randomizer $\calR_i$ to her neighbor list $\bma_i$. 
Then, $v_i$ changes the noisy neighbor list 
$\tbma_i \in \{0,1\}^n$ 
to a \textit{fake} neighbor list 
$\tbma_i^* \in \{0,1\}^n$ 
with the following probability:
\begin{align}
    \forall j \in [n]\setminus\{i\},~ \Pr(\ta_{i,j}^* = 1) = \begin{cases} 
\omega_1 & (\text{if } \ta_{i,j} = 1) \\ 
\omega_2 & (\text{otherwise}),
\end{cases} 
\label{eq:poisoning}
\end{align}
where $\ta_{i,j}$ and $\ta_{i,j}^*$ are the $j$-th elements of $\tbma_i$ and $\tbma_i^*$, respectively, 
and $\omega_1, \omega_2 \in [0,1]$. 
Finally, the malicious user $v_i$ copies $\tbma_i^*$ to $\tbma_i$ and sends $\tbma_i$ to the data collector, who calculates a noisy adjacency matrix $\tbmA$ from $\tbma_1, \cdots, \tbma_n$. 
We denote this attack by \Poison($\omega_1, \omega_2$).} 

\colorB{This attack is general and includes a lot of concrete attacks. 
For example, when $\omega_1 = \omega_2 = 1$, the malicious user $v_i$ always generates an all-ones vector $\tbma_i^* = (1,1,\ldots,1)$ (except for the $i$-th element $\ta_{i,i}^*=0$). 
When $\omega_1 = \omega_2 = 0$, $v_i$ generates an all-zeros vector $\tbma_i^* = (0,0,\ldots,0)$. 
When $\omega_1 = \omega_2 = 0.5$, $v_i$ generates a uniformly random $n$-dim binary vector as $\tbma_i^*$. 
Note that $v_i$ honestly sends $\tbma_i$ when $\omega_1 = 1$ and $\omega_2 = 0$. 
Thus, $v_i$ slightly changes $\tbma_i$ when $\omega_1$ and $\omega_2$ are close to $1$ and $0$, respectively.} 

\colorB{Since many social graphs are sparse in practice, the adversaries can degrade the classification accuracy of the GNN, especially when the fake neighbor list $\tbma_i^*$ is dense. 
Taking this into account, we evaluate \Poison($1, 1$) and \Poison($0.5, 0.5$) in our experiments. 
We refer to \Poison($1, 1$) and \Poison($0.5, 0.5$) as an \textit{all-ones attack} and \textit{random attack}, respectively.} 

\subsection{\colorB{Defenses}}
\label{sub:defenses}
\noindent{\textbf{\colorB{Existing Defense.}}}~~\colorB{Imola \textit{et al.} \cite{Imola_arXiv22_2} propose a defense against poisoning attacks for graph degree estimation. 
They focus on undirected graphs and use the fact that the adjacency matrix $\bmA$ is symmetric in this case. 
Specifically, their defense 
compares the $i$-th row of the noisy adjacency matrix $\tbmA$ with the $i$-th column of $\tbmA$. 
Then, it determines that user $v_i$ is malicious if the number of inconsistent elements is larger than a threshold.} 

\colorB{Unfortunately, 
it is difficult to apply their defense to our setting for two reasons. 
First, the defense in \cite{Imola_arXiv22_2} assumes that Warner's RR is used as a local randomizer $\calR_i$, making it easy to set a threshold so that the detection error probability is controlled. 
However, our DPRR applies edge sampling after Warner's RR, and the sampling probability $q_i$ is different from user to user. 
Moreover, the data collector does not know the value of $q_i$\footnote{\colorB{It is also possible for our DPRR to output the sampling probability $q_i$, as it provides $\epsilon_1$-edge DP. However, it does not address the issue, because the adversaries can change the value of $q_i$ to an arbitrary value.}}. 
Thus, it is difficult to set a threshold to control the detection error probability. 
Second, the defense in \cite{Imola_arXiv22_2} cannot be applied to directed graphs, as the adjacency matrix $\bmA$ is asymmetric in this case. 
Since we consider both directed and undirected graphs, a new defense is needed.} 

\smallskip
\noindent{\textbf{\colorB{Our Defense.}}}~~\colorB{We introduce a new defense that is applicable to both directed and undirected graphs. 
Our defense is based on the observation that the adversaries can degrade the accuracy of the GNN, especially when the fake neighbor list is dense (as described in Section~\ref{sub:defenses}).} 

\colorB{Specifically, 
recall that $\td_i$ ($= ||\tbma_i||_1$) is the number of 1s in the noisy neighbor list $\tbma_i$. 
In addition, recall that the parameter $p$ in Warner's RR is given by $p = \frac{e^{\epsilon_2}}{e^{\epsilon_2} + 1}$. 
In our defense, the data collector determines that user $v_i$ is \textit{malicious} if 
\begin{align*}
\td_i \geq \tau,
\end{align*}
where $\tau$ is a threshold given by 
\begin{align}
\tau = (n-1)p \left(1 + \frac{\log\frac{1}{\theta} + \sqrt{(\log\frac{1}{\theta})^2 + 8(n-1)p \log\frac{1}{\theta}}}{2(n-1)p} \right)
\label{eq:tau}
\end{align}
and $\theta \in [0,1]$ is a required value for the \textit{false positive probability} (i.e., the probability that an honest user is misclassified as a malicious user).} 

\colorB{The data collector discards noisy neighbor lists of users detected as malicious and uses only noisy neighbor lists of the remaining users (i.e., only a noisy graph composed of the remaining users). 
We denote this defense by \Defense($\theta$).
\begin{proposition}
\label{thm:defense}
The false positive probability of \Defense($\theta$) is smaller than or equal to $\theta$; i.e., for any honest user $v_i \in V$, we have 
\begin{align*}
\Pr(\td_i \geq \tau) \leq \theta,
\end{align*}
where $\td_i$ is the number of 1s in the noisy neighbor list $\tbma_i$, and $\tau$ is a threshold given by (\ref{eq:tau}). 
\end{proposition}
The proof is given in Appendix~\ref{sec:proof_defense}. 
In our experiments, we set $\theta = 0.05$.} 

\subsection{\colorB{Evaluation}}
\label{sub:evaluation}
\noindent{\textbf{\colorB{Experimental Set-up.}}}~~\colorB{We evaluated 
the data poisoning attacks and the defense in Sections~\ref{sub:data_poisoning} and \ref{sub:defenses}, respectively, 
using the three datasets in Section~\ref{sec:exp}. 
Specifically, we set the privacy budget $\epsilon$ to $\epsilon=1$ and the proportion $\lambda$ of non-private users to $\lambda=0$ (i.e., common setting). 
We set the proportion $\beta$ of malicious users to $\beta = 0$, $0.1$, $0.2$, $0.3$, $0.4$, or $0.5$.} 

\colorB{For attack algorithms, we evaluated \Poison($1, 1$) (i.e., all-ones attack) and \Poison($0.5, 0.5$) (i.e., random attack). 
In both \Poison($1, 1$) and \Poison($0.5, 0.5$), $\beta n$ malicious users change their noisy neighbor lists in the training graphs. 
Note that they leave their noisy neighbor lists in the testing graphs unchanged so that the difference between the training and testing graphs is large. 
We also confirmed that 
this attack results in lower accuracy than the attack that changes noisy neighbor lists in both the training and testing graphs.} 

\colorB{For a defense algorithm, we evaluated \Defense($0.05$). 
We used our DPRR with our privacy budget allocation method as a local randomizer, and evaluated the accuracy of this local randomizer with or without \Defense($0.05$). 
The other experimental settings are the same as Section~\ref{sub:setup}.}

\begin{figure}[t]
    \centering
    {\small (a) REDDIT-MULTI-5K (left: \Poison($1, 1$), right: \Poison($0.5, 0.5$))}\\
  \begin{minipage}{0.35\linewidth}
    \includegraphics[width=\linewidth]{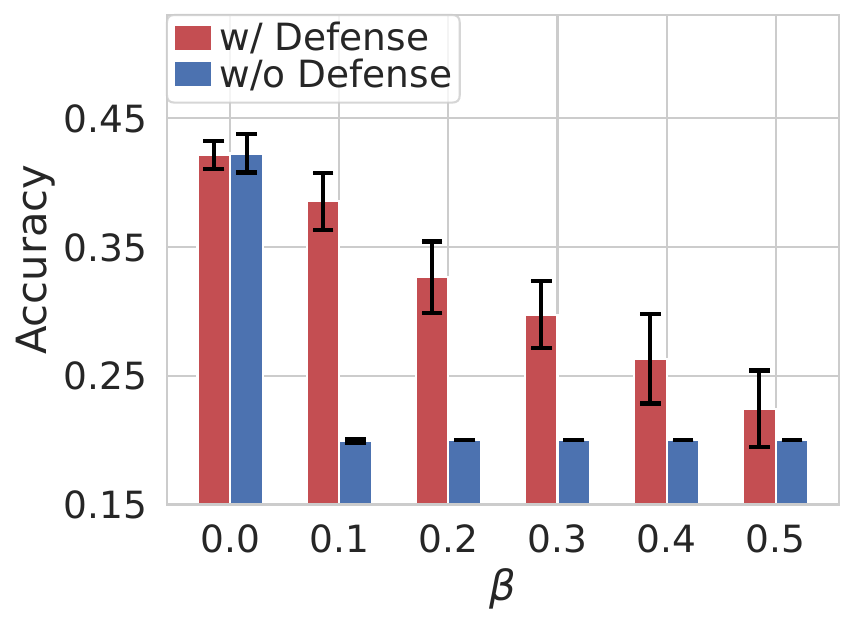}
  \end{minipage}
  \begin{minipage}{0.35\linewidth}
    \includegraphics[width=\linewidth]{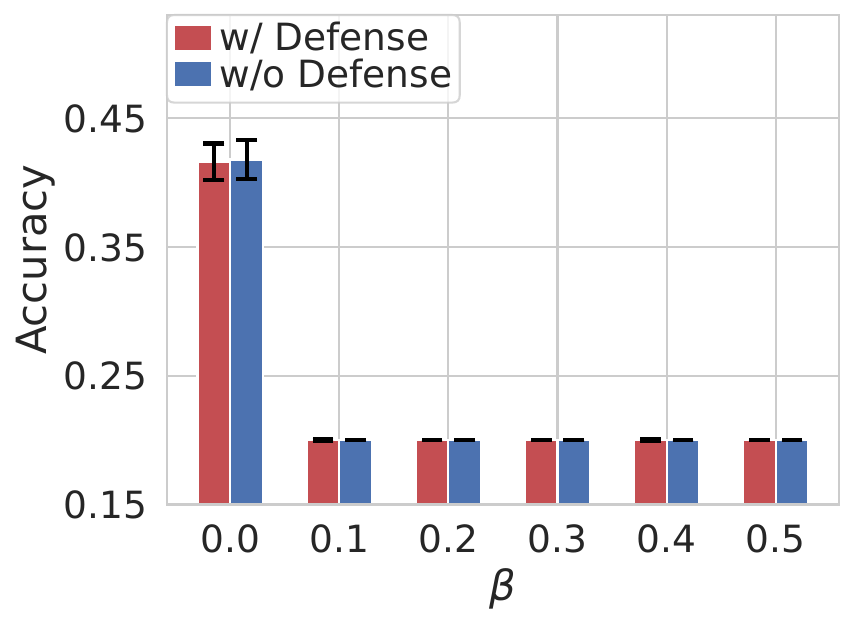}
  \end{minipage}\\
    {\small (b) REDDIT-BINARY (left: \Poison($1, 1$), right: \Poison($0.5, 0.5$))}\\
  \begin{minipage}{0.35\linewidth}
    \includegraphics[width=\linewidth]{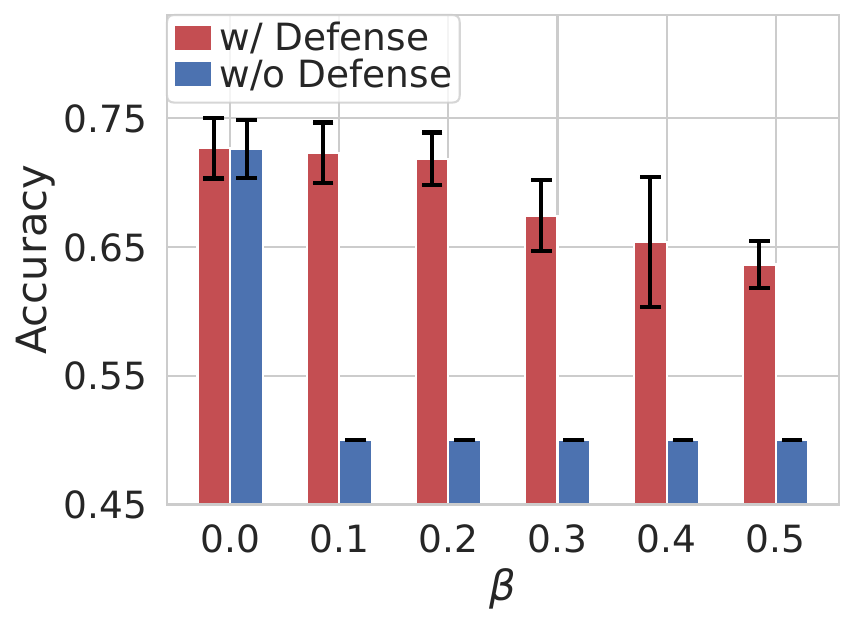}
  \end{minipage}
  \begin{minipage}{0.35\linewidth}
    \includegraphics[width=\linewidth]{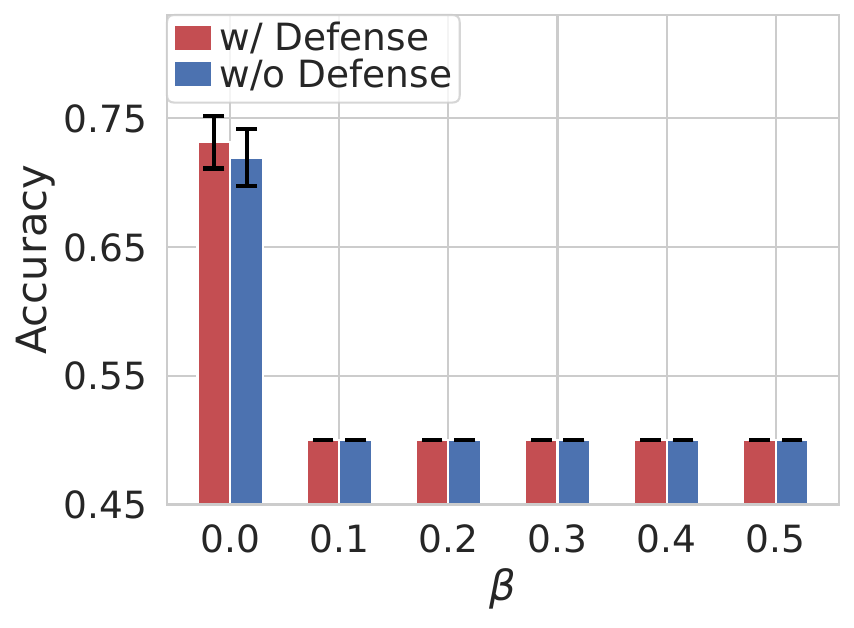}
  \end{minipage}\\
    {\small (c) Github StarGazers (left: \Poison($1, 1$), right: \Poison($0.5, 0.5$))}\\
  \begin{minipage}{0.35\linewidth}
    \includegraphics[width=\linewidth]{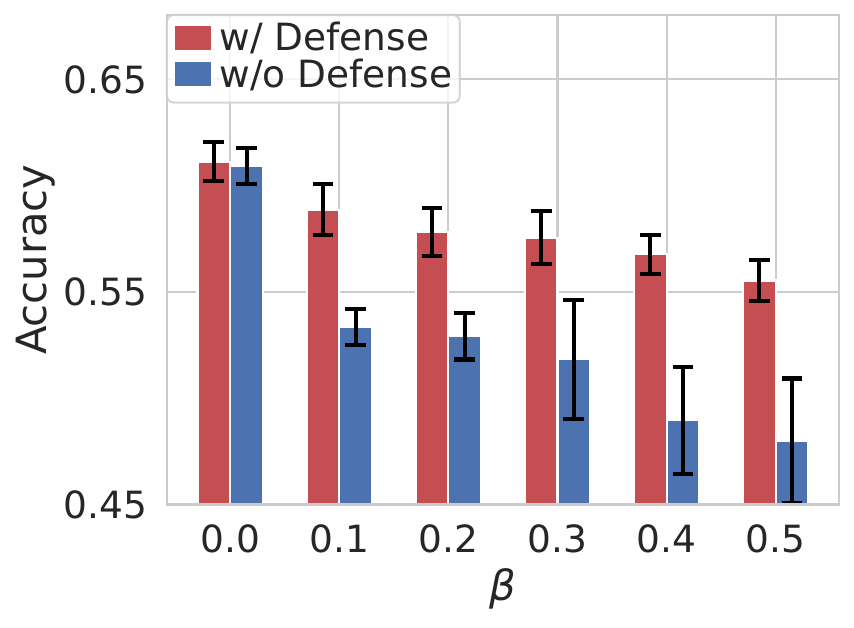}
  \end{minipage}
  \begin{minipage}{0.35\linewidth}
    \includegraphics[width=\linewidth]{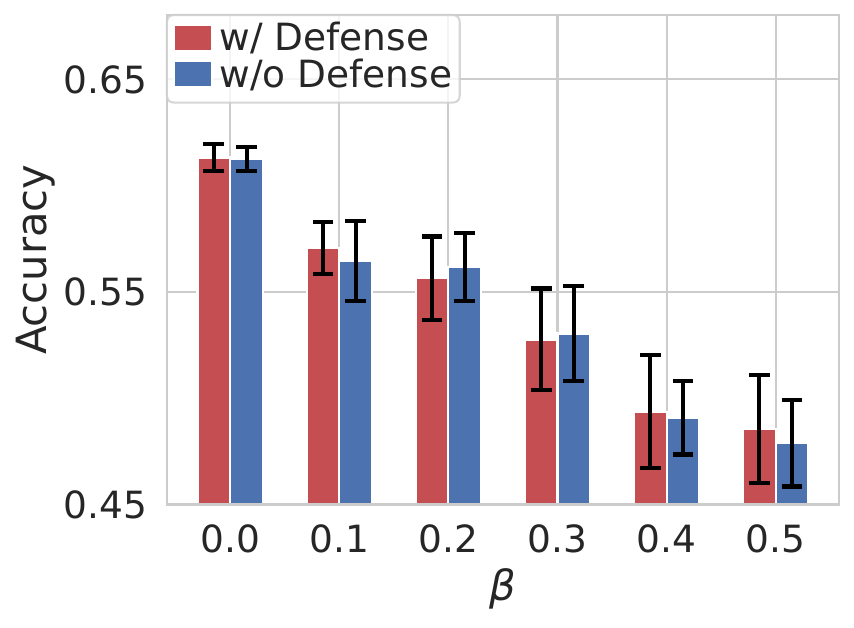}
  \end{minipage}
  \vspace{-2mm}
  \caption{\colorB{Classification accuracy for different proportions $\beta$ of malicious users ($\epsilon=1$, $\lambda=0$). 
  ``Defense'' represents \Defense($0.05$). 
  An error bar represents the standard deviation.}}
  \label{fig:accuracy_poisoning}
\end{figure}

\smallskip
\noindent{\textbf{\colorB{Experimental Results.}}}~~\colorB{Figure~\ref{fig:accuracy_poisoning} shows the results. 
The left figures show that the accuracy is significantly degraded by \Poison($1, 1$) when we do not introduce a defense. 
This attack is mitigated by using \Defense($0.05$). 
For example, when there is no attack ($\beta=0$), the accuracy of our DPRR is $0.61$ in Github StarGazers. 
When $\beta=0.5$, the accuracy of our DPRR with and without \Defense($0.05$) is $0.56$ and $0.48$, respectively, which means that \Defense($0.05$) effectively defends against \Poison($1, 1$).} 

\colorB{However, the right figures show that \Defense($0.05$) does not mitigate \Poison($0.5, 0.5$). 
For example, when $\beta=0.5$, the accuracy of our DPRR in Github StarGazers is about $0.48$, irrespective of the presence or absence of \Defense($0.05$). 
This means that \Defense($0.05$) is not effective for \Poison($0.5, 0.5$).} 

\smallskip
\noindent{\textbf{\colorB{Discussions.}}}~~\colorB{The reason why \Defense($0.05$) effectively defends against \Poison($1, 1$) but not against \Poison($0.5, 0.5$) lies in the threshold $\tau$ in (\ref{eq:tau}). 
In \Defense($0.05$), we set $\theta=0.05$ so that the false positive probability  (i.e., the probability that an honest user is misclassified as a malicious user) is smaller than or equal to $0.05$. 
As a result, the threshold $\tau$ becomes large. 
For example, when $n=100$ and $\epsilon_2=0.9$, the threshold $\tau$ in (\ref{eq:tau}) is: $\tau = 92.5$. 
Thus, \Defense($0.05$) can detect the all-ones attack \Poison($1, 1$) that always results in $\td_i = 99$ but cannot detect the random attack \Poison($0.5, 0.5$) that results in $\td_i = 49.5$ on average.}

\colorB{In summary, although \Defense($\theta$) in Section~\ref{sub:defenses} can be easily applied to both directed and undirected graphs, it cannot defend against all possible attacks. As described in Section~\ref{sub:defenses}, the existing defense \cite{Imola_arXiv22_2} cannot be applied to our setting, as (i) it assumes Warner's RR for a local randomizer $\calR$ and (ii) it cannot be applied to directed graphs. 
Our experimental evaluation shows that a new defense is needed to prevent all possible attacks more effectively. 
Developing such a new defense is left for future work.}

\section{Conclusion}
\label{sec:conclusion}
In this paper, we proposed the DPRR and a privacy budget allocation method to provide high accuracy in GNNs with a small privacy budget $\epsilon$ in edge LDP. 
Through experimental evaluation, 
we showed that the DPRR outperforms the three baselines (RR, \LocalLap{}, and \NPPartial{}) in terms of accuracy. 
We also showed that the DPRR is much more efficient than the RR in that it needs much less time for training and classification and much less memory. 
\colorB{We also evaluated the robustness of our DPRR against data poisoning attacks.}

Although we have focused on unattributed graphs, we can also provide LDP for both edges and feature vectors by combining our DPRR with the algorithm in \cite{Sajadmanesh_CCS21}. 
Specifically, the authors in \cite{Sajadmanesh_CCS21} propose an algorithm providing LDP for only feature vectors. 
They assume that original edges are public and use the original edges as input to their algorithms (Algorithms 2 and 3 in \cite{Sajadmanesh_CCS21}). 
Here, we can use noisy edges output by our DPRR as input to their algorithms. 
In other words, we can combine our DPRR with the algorithms in \cite{Sajadmanesh_CCS21} by replacing the original edges with the noisy edges output by our DPRR. 
Since the noisy edges provide LDP, we can provide LDP for both feature vectors and edges. 
We also note that although the algorithms in \cite{Sajadmanesh_CCS21} focus on node classification, they can be easily applied to graph classification by using the mean readout as a graph pooling method, as in our experiments. 
For future work, we would like to evaluate the accuracy of the combined algorithms using attributed graphs.

\section*{Acknowledgements}
This study was supported in part by JSPS KAKENHI 22H00521.

\bibliographystyle{plain}
\bibliography{main}

\appendix
\section{Proof of Proposition~\ref{thm:DPRR_privacy}}
\label{sec:proof_privacy}
Adding or removing one bit of $\bma_i$ will change a degree $d_i$ of user $v_i$ by one. 
Thus, the global sensitivity of the degree is $1$, and adding $\Lap(\frac{1}{\epsilon_1})$ to $d_i$ (line 2 in Algorithm~\ref{thm:DPRR_privacy}) provides $\epsilon_1$-edge LDP. 
The subsequent sampling probability tuning (lines 4-5) is a post-processing on the noisy degree $d_i^*$ ($=d_i + \Lap(\frac{1}{\epsilon_1}))$. 
In addition, Warner's RR with the flipping probability $1-p = \frac{1}{e^{\epsilon_2}+1}$ (line 6) provides $\epsilon_2$-edge LDP, as described in Section~\ref{sub:LDP}. 
The subsequent edge sampling (line 7) is a post-processing on the noisy neighbor list $\tbma_i$. 

Finally, we use the (general) sequential composition \cite{DP_Li} of edge LDP, which is proved in \cite{Imola_USENIX22}: 

\begin{lemma}[Sequential composition of edge LDP \cite{Imola_USENIX22}]
\label{lem:seqcomp}
For $i \in [n]$, let $\calR_i^1$ be a local randomizer of user $v_i$ 
that takes $\bma_i \in \{0,1\}^n$ as input. 
Let $\calR_i^2$ be a local randomizer of $v_i$ that depends on the output $\calR_i^1(\bma_i)$ of $\calR_i^1$. 
If $\calR_i^1$ provides $\epsilon_1$-edge LDP and for any $\calR_i^1(\bma_i)$, $\calR_i^2(\calR_i^1(\bma_i))$ provides $\epsilon_2$-edge LDP, then the sequential composition $(\calR_i^1(\bma_i), \calR_i^2(\calR_i^1(\bma_i))(\bma_i))$ provides $(\epsilon_1 + \epsilon_2)$-edge LDP. 
\end{lemma}
In our case, $\calR_i^1$ is the Laplacian mechanism followed by the sampling probability tuning, and $\calR_i^2$ is Warner's RR followed by the edge sampling. 
$\calR_i^2$ depends on the output $q_i$ of $\calR_i^1$. 
Thus, by 
Lemma~\ref{lem:seqcomp} (and the post-processing invariance), the DPRR provides $(\epsilon_1 + \epsilon_2)$-edge LDP. \qed

\section{\colorB{Proof of Proposition~\ref{thm:defense}}}
\label{sec:proof_defense}
\colorB{Assume that user $v_i$ is honest. 
The number $\td_i$ ($= ||\tbma_i||_1$) of 1s in the noisy neighbor list $\tbma_i$ is maximized when the original neighbor list $\bma_i$ is an all-ones vector $\bma_i = (1,1,\ldots,1)$ (except for the $i$-th element $a_{i,i}=0$) and the sampling probability $q_i$ is $q_i = 1$. 
In this case, each element of the noisy neighbor list $\tbma_i$ (except for the $i$-th one) follows the Bernoulli distribution with parameter $p = \frac{e^{\epsilon_2}}{e^{\epsilon_2} + 1}$. 
Thus, we can use the multiplicative Chernoff bound \cite{probability_computing} as follows:
\begin{align}
\Pr(\td_i \geq (1 + \delta)\mu) \leq e^{- \frac{\delta^2 \mu}{2 + \delta}},
\label{eq:Pr_td_i}
\end{align}
where $\mu = (n-1)p$ and $\delta \geq 0$.}

\colorB{Let $\theta = e^{- \frac{\delta^2 \mu}{2 + \delta}}$. Then, we have
\begin{align}
\theta = e^{- \frac{\delta^2 \mu}{2 + \delta}} 
&\iff \log\frac{1}{\theta} = \frac{\delta^2 \mu}{2 + \delta} \nonumber\\
&\iff \mu \delta^2 - \left(\log\frac{1}{\theta}\right) \delta - 2 \left(\log\frac{1}{\theta}\right) = 0 \nonumber\\
&\iff \delta = \frac{\log\frac{1}{\theta} + \sqrt{(\log\frac{1}{\theta})^2 + 8\mu \log\frac{1}{\theta}}}{2\mu} \hspace{5mm} \text{(as $\delta \geq 0$)}.
\label{eq:theta_delta}
\end{align}
By (\ref{eq:Pr_td_i}) and (\ref{eq:theta_delta}), we have 
\begin{align*}
\Pr(\td_i \geq \tau) \leq \theta,
\end{align*}
where $\tau$ is given by (\ref{eq:tau}). \qed
}

\end{document}